\documentclass[11pt]{article}
\pdfoutput=1
\usepackage{jcappub,amsmath,amssymb}
\usepackage{enumerate}
\usepackage{bm}
\usepackage{bbm}
\usepackage[usenames,dvipsnames]{xcolor}
\usepackage{graphics}
\usepackage{tikz,todonotes}
\usepackage[utf8]{inputenc}
\usetikzlibrary{arrows,positioning,decorations.markings,decorations.pathmorphing,calc}

\definecolor{hyperref}{RGB}{026,028,087}

\def\gsim{ \lower .75ex \hbox{$\sim$} \llap{\raise .27ex \hbox{$>$}} }
\def\lsim{ \lower .75ex \hbox{$\sim$} \llap{\raise .27ex \hbox{$<$}} }
\def\be{\begin{equation}}
\def\ee{\end{equation}}
\def\bea{\begin{eqnarray}}
\def\eea{\end{eqnarray}}

\newcommand{\ba}{\begin{array}}
\newcommand{\ea}{\end{array}}

\newcommand{\commentout}[1]{}

\newcommand{\pa}{\partial}

\newcommand{\cS}{{\cal{S}}}
\newcommand{\cI}{{\cal{I}}}

\newcommand{\comment}[1]{}

\newcommand{\bs}{\begin{split}}

\newcommand{\Ostro}{{Ostrogradski}}
\newcommand{\dl}{{\it DL}}
\newcommand{\eom}{{\it eom}}
\newcommand{\eoms}{{\it eoms}}
\newcommand{\dof}{{\it dof}}
\newcommand{\dofs}{{\it dofs}}
\newcommand{\eft}{{\it EFT}}
\newcommand{\St}{{St\"uckelberg}}
\newcommand{\Sting}{{St\"uckelberging}}

\newcommand{\good}{{ \Large \color{ForestGreen} $\checkmark$}} 
\newcommand{\bad}{{\color{red} \sffamily X}}

\def\ba{\begin{eqnarray}}
\def\ea{\end{eqnarray}}

\def\nn{\nonumber}
\def\ni{\noindent}

\def\({\left(}
\def\){\right)}

\newcommand*{\mathcolor}{}
\def\mathcolor#1#{\mathcoloraux{#1}}
\newcommand*{\mathcoloraux}[3]{%
  \protect\leavevmode
  \begingroup
    \color#1{#2}#3%
  \endgroup
}
\newlength{\stheight}
\newcommand\textst[1][fu-grey]{
	\ifmmode\setlength{\stheight}{+1.0ex}
	\else\setlength{\stheight}{+0.5ex}
	\fi
	\bgroup\markoverwith{\textcolor{#1}{\rule[\the\stheight]{2pt}{1.0pt}}}\ULon
} 
\newcommand{\textins}[2][fu-grey]{
	\ifmmode\mathcolor{#1}{#2}
	\else\textcolor{#1}{#2}\@\,
	\fi
}

\newcommand{\w}{\wedge}

\newcommand{\N}{{\cal N}}

\def\({\left(}
\def\){\right)}

\notoc

  \tikzstyle{vecArrow} = [thick, decoration={markings,mark=at position
   1 with {\arrow[semithick]{open triangle 60}}},
   double distance=1.4pt, shorten >= 5.5pt,
   preaction = {decorate},
   postaction = {draw,line width=1.4pt, white,shorten >= 4.5pt}]

\begin{document}

\title{Generalised matter couplings \\in massive bigravity}

\author[a,b]{Scott Melville,}
\author[b,c]{Johannes Noller}

\affiliation[a]{Harvard University, Cambridge, Massachusetts 02138, USA}
\affiliation[b]{The Queen's College, High Street, Oxford, OX1 4AW, UK}
\affiliation[c]{Astrophysics, University of Oxford, DWB, Keble Road, Oxford, OX1 3RH, UK} 

\emailAdd{scott.melville@oxon.org}
\emailAdd{noller@physics.ox.ac.uk}

\abstract{
We investigate matter couplings in massive bigravity. We find a new family of such consistent couplings,  including and extending known consistent matter couplings, and we investigate their decoupling limits, ADM decompositions, Higuchi bounds and further aspects. We show that differences to previous known consistent couplings only arise beyond the $\Lambda_3$ decoupling limit and discuss the uniqueness of consistent matter couplings and how this is related to the so-called symmetric vielbein condition. Since we work in a vielbein formulation, these results easily generalise to multi-gravity.   
}


\maketitle
\newpage

\setcounter{tocdepth}{2}
\tableofcontents

\section{Introduction} 
\label{sec1:intro}

Interest in massive gravity has experienced a remarkable resurgence in recent years. Reasons for this are manifold, but here we emphasise two: Firstly the cosmologically motivated hope that such theories will have something to say about the (late-time) accelerated expansion of the universe, providing a satisfactory model of `dark energy'/modified gravity. Secondly the theoretically motivated desire to understand the complete space of consistent spin-2 theories. With a spin-2 field, or a graviton, being the quintessential gravitational degree of freedom (\dof), understanding what possible consistent interactions and theories one can build for such a field promises to teach us much about gravity.

In this paper we will focus on the second motivation for studying massive gravity and related theories. Until roughly five years ago, General Relativity (GR) -- the unique consistent classical theory of a massless spin-2 field -- was widely believed to be the only classically consistent spin-2 theory. However, following the initial discovery of a set of consistent massive spin-2 theories -- so-called ghost-free massive gravity \cite{deRham:2010ik, deRham:2010kj, Hassan:2011vm} and bigravity \cite{Hassan:2011hr, Hassan:2011zd} -- much progress has been made in understanding the landscape of consistent spin-2 field theories. 
In this paper we focus on one aspect of this landscape: new consistent and non-minimal couplings to matter. The known consistent non-derivative such couplings are those proposed by \cite{deRham:2014naa,Noller:2014sta}.\footnote{Derivative matter couplings have recently been proposed in \cite{Heisenberg:2015wja}.}
%
These couplings have since been further investigated in \cite{deRham:2014fha,Schmidt-May:2014xla,Solomon:2014iwa,Gao:2014xaa,Hinterbichler:2015yaa,deRham:2015cha,Huang:2015yga,Blanchet:2015bia,Heisenberg:2015iqa,Matas:2015qxa,Enander:2014xga,Gumrukcuoglu:2014xba,Comelli:2015pua,Gumrukcuoglu:2015nua,DeFelice:2015yha,Lagos:2015sya}, in particular their cosmological solutions have been explored in \cite{Solomon:2014iwa,Enander:2014xga,Gumrukcuoglu:2014xba,Comelli:2015pua,Gumrukcuoglu:2015nua,Lagos:2015sya}.   
Working in the vielbein formulation, here we show that new consistent matter couplings can be found, with differences to the known couplings of \cite{deRham:2014naa,Noller:2014sta} manifesting themselves beyond the $\Lambda_3$ decoupling limit (\dl) or, somewhat equivalently as we will see, when the symmetric vielbein condition does not hold. 
\\

\ni {\it Outline}: 
The outline for this paper is as follows. In section \ref{sec2:cand} we introduce possible candidate matter couplings in the vielbein formulation of bigravity (which we also briefly review). In the following four sections we subject these candidates to several tests: in section \ref{sec3:matterloops} we check whether the pure spin-2 interactions generated by the matter coupling (through the presence of an effective cosmological constant, e.g. generated via matter loops) are stable for the candidate couplings; in section \ref{sec4:constraint} we perform mini-superspace and full ADM analyses to check for the presence of additional propagating (ghost-like) \dof; in section \ref{sec5:decoupling} we then perform a decoupling limit analysis, investigating whether such couplings contribute to a valid low-energy effective field theory (\eft) (even where the full ADM analysis revealed a ghost, which may propagate outside the \eft's regime of validity); in section \ref{sec6:higuchi} we check for the potential presence of Higuchi ghosts and derive the associated Higuchi bounds on the theory. Finally, we discuss the uniqueness of our new consistent matter couplings in section \ref{sec7:unique}, before concluding in section \ref{sec:conc} and collecting and discussing further computations in the appendices.
\\

\ni {\it Conventions}: Throughout this paper we use the following conventions. $D$ refers to the number of spacetime dimensions, frequently taken to be four for simplicity, and we use Greek letters $\mu, \nu, \ldots$ and lower case Latin letters $a, b, c, \ldots$ to denote spacetime indices, which are raised and lowered as specified (this issue is not trivial in theories with several vielbeins/spin-2 fields/`metrics').  Capital Latin letters $A$, $B,\ldots$ are reserved for Lorentz indices and are raised and lowered with the Minkowski metric $\eta_{AB}$.  Bracketed indices $(i),(j),\ldots$, label the different vielbeins/spin-2 fields -- label indices are not automatically summed over and whether they are upper or lower indices carries no meaning. We denote the completely anti-symmetric epsilon symbol by $\tilde \epsilon$ and define it such that $\tilde\epsilon_{012\cdots D}=1$ regardless of the signature of the metric or the position (up/down) of indices (hence $\tilde\epsilon^{012\cdots D}=\tilde\epsilon_{012\cdots D}=1$).  
\\

\section{Candidate couplings to matter}\label{sec2:cand}

In this section we lay out the structure of the couplings between spin-2 fields and matter fields that we will consider. Our construction aims to find the general form of allowed couplings in the vielbein formulation.

\subsection{The vielbein picture for massive and bi-gravity}

{\bf Vielbeins}: Throughout this paper we will primarily be working in the vielbein formulation for gravity. Corresponding to each spin-2 field/metric $g_{(i)}$ and its inverse $g^{-1}_{(i)}$ we have a vielbein $E_{(i)}$ and an inverse vielbein $E^{-1}_{(i)}$ satisfying 
\begin{align}
g_{\mu\nu}^{(i)} &= E_{(i)}{}_{\mu}^{\ A} E_{(i)}{}_{\nu}^{\ B} \eta_{AB},  &g^{-1}_{(i)}{}^{\mu\nu} &= E^{-1}_{(i)}{}^{\mu}_{\ A} E^{-1}_{(i)}{}^{\nu}_{\ B} \eta_{AB},
\end{align}
where we emphasise that the vielbein is not in general symmetric in its two indices and that it satisfies the following conditions
\begin{align}
\label{cond}
E^{-1}_{(i)}{}^\mu_{\ A} E_{(i)}{}_{\nu}^{\ A} &= \delta^\mu_\nu, &E^{-1}_{(i)}{}^\mu_{\ A} E_{(i)}{}_{\mu}^{\ B} &= \delta^B_A.
\end{align}
Lorentz indices (capital Latin letters) are raised/lowered with the flat Minkowski metric $\eta_{AB}$ and spacetime indices (lower case Greek indices) on $E_{(i)}$ are raised/lowered with the full metric $g_{(i)}$. However, since there are in principle several distinct spacetime `metrics' in the theory, here we choose the convention to raise/lower spacetime indices with the flat spacetime metric $\eta_{\mu\nu}$ and to explicitly write out any metric $g_{(i)}$ whenever it is used to raise/lower spacetime indices instead.
\\

\noindent {\bf Massive and Bi-gravity}: 
The known ghost-free potential interactions for $\N$ spin-2 fields are those of ghost-free massive gravity \cite{deRham:2010ik,deRham:2010kj,Hassan:2011hr}, Bigravity \cite{Hassan:2011tf,Hassan:2011zd,Hassan:2011ea} and Multi-Gravity \cite{Hinterbichler:2012cn}. In terms of vielbeins they can all be cast in the unified format \cite{Hinterbichler:2012cn} (in $D$ dimensions)
 \bea   \label{density}
 {\cI}_{(i_1 i_2 \ldots i_D)}
 \equiv  \tilde \epsilon_{A_1 A_2 \cdots A_D} \, \tilde  \epsilon^{\mu_1 \mu_2 \cdots \mu_D} \, 
 {E}_{(i_1)}{}_{\mu_1}^{\ A_1} \,  {E}_{(i_2)}{}_{\mu_2}^{\ A_2} \cdots {E}_{(i_D)}{}_{\mu_D}^{\ A_D},
\eea
where the indices $(i_1 i_2 \ldots i_D)$ keep track of which fields are interacting.\footnote{Equivalently one can write this in terms of vielbein one-forms ${\bf E}_{(i)}^A=E_{(i)}{}_\mu^{\ A}dx_{(i)}^\mu$. The interaction terms can then be written 
\bea \label{genint}
{\cI}_{(i_1 i_2 \ldots i_D)} d^D x &\equiv & \tilde  \epsilon_{A_1 A_2 \cdots A_D}\, {\bf E}^{A_1}_{(i_1)}\w {\bf E}^{A_2}_{(i_2)}\w \ldots\w {\bf E}^{A_D}_{(i_D)},
\eea
in terms of the usual wedge product. The determinant-like nature of the interaction terms ensures that the order of labels in \eqref{density} $(i_1 i_2 \ldots i_D)$ is irrelevant.} 
Ghost-free massive gravity potential interactions then consist of all the ways to build \eqref{density} with a single dynamical vielbein $E_{(1)}$ and a non-dynamical reference vielbein, which is held fixed (in the case of a flat reference metric this non-dynamical vielbein is $E_{(0)}{}_\mu^{\ A} = \delta_\mu^{\ A}$). Ghost-free Bigravity consists of all interactions \eqref{density} that can be built with two dynamical vielbeins $E_{(1)}$ and $E_{(2)}$, and so on. The most general known, fully ghost-free potential interaction for $\N$ spin-2 fields and in $D$ dimensions can therefore be written as
\bea \label{potential}
{\cal S}_{\rm pot} = \sum_{i_j}^{\N} m^2_{(i_1 i_2 \ldots i_D)} M_{\rm Pl}^{D-2} \int d^Dx  \; c_{(i_1 i_2 \ldots i_D)} {\cI}_{(i_1 i_2 \ldots i_D)},
\eea
where $m^2_{(i_1 i_2 \ldots i_D)} M_{\rm Pl}^{D-2}$ is the coupling constant for a given interaction term and the $c_{(i_1 i_2 \ldots i_D)}$ and $m^2_{(i_1 i_2 \ldots i_D)}$ are constant coefficients completely symmetric in all the $i_j$.

\subsection{Possible effective (matter) vielbeins} 


Throughout this paper we consider theories of the form
\be \label{S-gen}
\cS = \sum_{i=1}^{\N} {M_{(i)}^{D-2}\over 2} \int d^Dx\ \det(E_{(i)}) R[E_{(i)}] + {\cal S}_{\rm pot} + {\cal S}_{\rm mat}[g_{\rm eff},\Phi_i],
\ee
i.e. theories of multiple spin-2 fields in the vielbein formulation, whose kinetic sector is a superposition of Einstein-Hilbert terms and that have potential interactions of the type \eqref{potential}. The coupling to matter fields $\Phi_i$ occurs in accordance with the weak equivalence principle, i.e. matter uniformly and minimally couples to a single effective matter metric that is a function of the vielbeins and inverse vielbeins living in the theory.\footnote{It was recently and explicitly shown in \cite{Matas:2015qxa} that this assumption does not need to be imposed separately for the construction of consistent matter couplings. In an \eft{}-sense these can of course be constructed just from the requirement of having ghost-free decoupling limit interactions. In other words, we can couple some matter consistently to one effective metric/vielbein as discussed here and some other matter consistently to another effective metric/vielbein as discussed here. Such couplings will trivially violate the weak equivalence principle, but are consistent (at least in the decoupling limit), since we have coupled matter to (different) consistent effective metrics/vielbeins of a ghost-free form.} More specifically we write
\be
g^{\rm eff}_{\mu\nu} = \tilde E_\mu^A \tilde E_\nu^B \eta_{AB},
\ee
where the effective matter vielbein $\tilde E$ will be a general function of the vielbeins and inverse vielbeins in the theory. 
\\

\noindent {\bf Effective vielbeins}:  We will assume that the effective matter vielbein can be expressed as a power-law expansion in all the fundamental vielbeins (and their inverses) in the theory.\footnote{Somewhat equivalently, taking (the components of) the effective matter metric $g_{\rm eff}$ to be real analytic functions of (the components of) the metrics/vielbeins/spin-2 fields in the theory, we are guaranteed a locally convergent series expansion of $\tilde{E}$ in powers of $E_{(i)}$. So our above assumption of having a meaningful expansion of the effective vielbein in terms of powers of the fundamental vielbeins and their inverses is effectively assuming analyticity of the effective vielbein/metric.} 
The effective vielbein $\tilde{E}$ must have one free spacetime and one free Lorentz index, so we write the possible building blocks of $\tilde{E}$ order-by-order as contractions of vielbeins and inverse vielbeins (labelled by a `species' index $(i)$) as 
\begin{align}
\nn \mathsf{1-vielbein} \; &: \;\;  {E_{(p)}}_{\mu}^{\; A} \\ 
\nn \mathsf{2-vielbein} \; &: \;\;\mathsf{ None \; possible } \\
\nn \mathsf{3-vielbein} \; &: \;\;  {E_{(p)}}_{\mu}^{\; B} {E_{(q)}}_{\lambda}^{\; C} {E_{(p)}^{-1}}^{\lambda}_{\; D} \eta_{BC} \, \eta^{DA}   \\
\nn \mathsf{4-vielbein} \; &: \;\;\mathsf{ None \; possible } \\
\mathsf{5-vielbein} \; &: \;\;  {E_{(p)}}_{\mu}^{\;E} {E_{(q)}^{-1}}^{\rho}_{\; E}  {E_{(p)}}_{\rho}^{\; B} {E_{(q)}}_{\lambda}^{\; C} {E_{(p)}^{-1}}^{\lambda}_{\; D} \eta_{BC} \, \eta^{DA} ,  \, \nn     \\ 
\vdots     \label{bbs}
\end{align}
Here we have suppressed any combinations that become Kronecker deltas or reduce to one of the combinations listed above by virtue of the relation \eqref{cond}. In general the effective vielbein can  be an arbitrary linear combination of the above building blocks, with scalar functions of the $E_{(i)}$ as prefactors. We discuss some additional details of the structure of these terms in appendix \ref{appendix:multi}.

As a concrete example consider the case when the effective vielbein simply is one of the building blocks in \eqref{bbs} (i.e. not a superposition). While no effective 2- or 4-vielbein couplings are possible, effective 3- and 5-vielbein couplings respectively lead to effective matter metrics
\bea
g^{\rm eff}_{\mu \nu} &=& \left( E_{(p)} E_{(q)} \right)_{\mu \lambda}  \left( E_{(p)}^{-1} E_{(p)}^{-1} \right)^{\lambda \rho} \left( E_{(p)} E_{(q)} \right)_{\nu \rho}  \\
g^{\rm eff}_{\mu \nu} &=& \left( E_{(p)} E_{(q)}^{-1}  \right)_\mu^{\;\;\; \lambda} \left( E_{(p)} E_{(q)}^{-1}  \right)_\nu^{\;\;\; \epsilon} 
\left( E_{(p)}  E_{(q)} \right)_{\lambda \rho} \left( E_{(p)}  E_{(q)} \right)_{\epsilon \delta}  \left( E_{(p)}^{-1} E_{(p)}^{-1}  \right)^{\rho \delta}  ,
\eea
and so on. Bracketed pairs of vielbeins/inverse vielbeins denote pairs whose Lorentz indices are contracted. We emphasise that contractions of vielbeins with inverse vielbeins do not reduce to Kronecker deltas, if the species index $(i)$ does not match.
\\

\ni {\bf Superpositions}: One may now build candidate effective vielbeins that are made up of superpositions of the vielbein building blocks \eqref{bbs}. For example, such a candidate coupling could be
\be \label{super1}
\tilde E_{\rm eff}{}_{\mu}^{D} = \alpha {E_{(1)}}_{\mu}^{\; A} {E_{(2)}}_{\lambda}^{\; B} {E_{(1)}^{-1}}^{\lambda}_{\; C} \eta_{AB}\eta^{CD} + \beta {E_{(2)}}_{\mu}^{\; A} {E_{(1)}}_{\lambda}^{\; B} {E_{(2)}^{-1}}^{\lambda}_{\; C} \eta_{AB} \eta^{CD},
\ee
where $\alpha$ and $\beta$ are arbitrary scalar functions of the $E_{(i)}$. For simplicity we will mostly consider single terms throughout, but we will return to superpositions in section \ref{sec5:decoupling} when discussing the decoupling limit and the symmetric vielbein condition. Note that, in the massive gravity case where one of the two vielbeins is non-dynamical, any superposition like \eqref{super1} will automatically reduce back to the massive gravity limit of the construction of \cite{deRham:2014naa,Noller:2014sta} (since the `Minkowski vielbein' is simply a Kronecker delta).

\begin{table}
\small
\begin{center}
\def\arraystretch{1.5}
\begin{tabular}{|c | c | c c c | c | c|}
\hline
 &  Vacuum   & MSS Boson & MSS Fermion & Full ADM	& $\Lambda_3$ DL & Higuchi  			 	   	\\
\hline
Rank-2 & \good  & \good  &  \good  &  \bad  (minimal) 	&   \good     & \good   		   		\\
Rank-0 &  \good  &  \good  &  \bad (conformal)    & \bad  (conformal) 	& \bad  (conformal)   &  \bad  (conformal)     	 			\\
Mixed &  \good  & \bad  & \bad  (tuned) & \bad		& \bad ($\overset{svc}{\to}$ rank-2)   &  \bad ($\overset{svc}{\to}$ rank-2) \\  \hline
\end{tabular}

\caption{The three constructions we consider (rank-2, rank-0, mixed) and where their ghosts appear. Exceptional cases are given in brackets, e.g. \bad (condition) means the construction fails this check, unless the given condition is satisfied. All constructions are built, such that they (or at least their individual components) give rise to healthy pure-spin 2 interactions (from an effective cosmological constant/one-loop (matter) corrections) in the matter sector. We also investigate these couplings in the mini-superspace (MSS) for matter actions containing scalar, Yang-Mills and Dirac fields, finding that only the rank-2 couplings can give ghost-free MSS couplings for all species simultaneously. Similarly, in the $\Lambda_3$ decoupling limit,
only rank-2 constructions are healthy for general $\mathcal{L}_M$. The mixed constructions are only healthy if they reduce to rank-2 couplings on imposing the symmetric vielbein condition (denoted {\it svc} in the table). 
All couplings have a ghost in the full ADM construction, which lies above the $\Lambda_3$ scale in the case of rank-2 constructions, however, and is hence not part of the healthy low-energy effective field theory valid (at least) up to the scale $\Lambda_3$. Finally we check if the Higuchi bound can be satisfied for the different couplings.
}
\label{tab:results}
\end{center}
\end{table}

\subsection{Metric representation and symmetric vielbein condition} 
\label{sec:symmvielbeincond}

At the level of kinetic (Einstein-Hilbert) terms, a multi-gravity theory with $N$ dynamical fields possesses $N$ copies of local Lorentz transformations $LLT_{(i)}$. These symmetries are generically broken down to their diagonal subgroups by the potential interactions ${\cal S}_{\rm pot}$ \eqref{potential} (for details see \cite{ArkaniHamed:2002sp,Hinterbichler:2012cn,Noller:2013yja}). They may be re-introduced at the expense of introducing a (gauge) \St{} field $\Lambda$ through the following replacement  
\begin{equation}
E_{(1)}{}_\mu^A \rightarrow \Lambda^A_B E_{(1)}{}^B_\mu.
\end{equation}  
Clearly the kinetic (Einstein-Hilbert) terms are gauge-invariant under this replacement and therefore do not contribute to the $\Lambda$ equations of motion. $\Lambda$ is then a non-dynamical, auxiliary field which may simply be integrated out. For bigravity, in the absence of coupling to matter fields $\Phi_i$, the $\Lambda$ \eom{} implies (in matrix notation)\footnote{Note that the vielbein formulation of massive, bi- and multi-gravity permits several branches and that the symmetric vielbein condition \eqref{eqn:svc} is only imposed in some of them \cite{Deffayet:2012zc,Banados:2013fda}. Here we implicitly assume that we are working in a branch of bigravity with minimal matter coupling where the $\Lambda$ \eom{} imposes the symmetric vielbein condition. To our knowledge it is still an open question whether these branches are disconnected or not. 
} \cite{Ondo:2013wka}
\begin{align}
{\cal E}_\Lambda = \frac{\delta S_{\text{pot}}}{\delta \Lambda} &= 0  &\Longrightarrow &  &(\Lambda E_{(1)})^T \eta E_{(2)} = E_{(2)}^{T} \eta (\Lambda E_{(1)}) .
\end{align}
This is known as the Deser-von-Nieuwenhuizen (DvN) condition and ensures that an equivalent metric formulation exists (where $S_{\text{pot}}$ may be written in terms of elementary symmetric polynomials). Significantly, this condition is independent of the coefficients $c_{(i_1,i_2...i_D)}$. 
The DvN condition allows us to choose a gauge in which the vielbeins commute. In particular this means we can choose (i.e. gauge fix using a copy of {\it LLT}) a $\Lambda$ such that
\be \label{eqn:svc}
E_{(1)}{}_\mu^{A} E_{(2)}{}_{\nu}^{B} \eta_{AB} = E_{(1)}{}_\nu^{A} E_{(2)}{}_{\mu}^{B} \eta_{AB}
\ee
If this commutativity holds for all the vielbeins in the theory, the vielbein strings in \eqref{bbs} may be freely reordered so that vielbeins cancel against inverse vielbeins and all building blocks in \eqref{bbs} simply collapse to single 1-vielbein strings. If the symmetric vielbein condition holds for all vielbeins, all the (superpositions of) vielbein building blocks \eqref{bbs} discussed in the previous subsection collapse to the couplings proposed by \cite{deRham:2014naa,Noller:2014sta}.

As should already be clear from the above, in general the DvN gauge condition is not imposed dynamically. When a coupling to matter fields $\Phi_i$ is introduced, we have another contribution to the Lorentz \eoms, 
\begin{equation}
\frac{\delta S_{\text{pot}}}{\delta \Lambda} + \frac{\delta S_{\text{mat}}}{\delta \Lambda} = 0.
\end{equation}
As pointed out in \cite{Noller:2014sta,Hinterbichler:2015yaa,deRham:2015cha}, this means we are no longer guaranteed that an equivalent metric formulation of the theory exists.
However, even when this is the case and the DvN condition therefore does not hold for the full theory, particular solutions of the theory (e.g. ones considering highly symmetric FLRW backgrounds) or particular (low-energy) scaling limits (e.g. the $\Lambda_3$ decoupling limit discussed in section \ref{sec5:decoupling}) can recover this condition. In these limits an equivalent metric formulation can therefore still exist. We will discuss these points further in what follows.

\section{Effective cosmological constants and matter loops} 
\label{sec3:matterloops}

\ni {\bf Healthy spin-2 interactions}: An important and very powerful check for any candidate coupling to matter is the requirement that any effective cosmological constant contribution it generates must be healthy.  By this we mean that matter loops should not break the healthy form of the pure spin-2 interactions at tree level and, since they will (also) generate a contribution to the overall action of the form
\begin{equation} \label{lambda}
S =  \int d^D x \, \det (\tilde{E}) \,  \tilde \Lambda ,
\end{equation}
we would like this to be of the ghost-free form \eqref{density}, where $\tilde \Lambda$ is a constant. This is a pure spin-2 interaction term by virtue of the ${\rm det}(\tilde E)$. Equivalently \eqref{lambda} can be seen as the zeroth order Lovelock invariant, or a cosmological constant, for $g_{\rm eff}$. There are a number of ways in which to achieve this. It will prove useful to divide the couplings which generate pure spin-2 interactions of the form \eqref{genint} into three classes:
\begin{itemize} 
  \item[(a)] {\bf Rank-2 construction}: The effective matter vielbein $\tilde E$ is a (superposition of) combinations of $N$ vielbeins and inverse vielbeins, with constant coefficients. There are $n_{(i)}$ copies of vielbein $E_{(i)}$  and $\bar n_{(i)}$ copies of its inverse involved in a single vielbein string. If the effective matter vielbein $\tilde E$ only consists of such a single string as listed in \eqref{bbs}, $\tilde E$ satisfies
\be \nn
\tilde E = \tilde E \left[E_{(1)},\ldots,E_{(N)}, E^{-1}_{(1)},\ldots,E^{-1}_{(N)}\right], 
\ee
\be
\exists!j: n_{(j)} - \bar n_{(j)} = 1, \quad \quad \quad \quad \forall i \neq j: n_{(i)} - \bar n_{(i)} = 0,    
 \label{eqn:rank2}
\ee
where $i,j \in \{ 1,\ldots,N \}$ and we have required that there exists one and only one vielbein $j$ that satisfies $n_{(j)} - \bar n_{(j)} = 1$. In the vielbein language a coupling of this kind was recently introduced in \cite{Noller:2014sta}, where the effective vielbein $\tilde{E}$ was constructed from vielbeins only (as a linear superposition of 1-vielbein strings). If the symmetric vielbein condition holds, all the rank-2 couplings reduce to  such a coupling $\tilde{E} = \sum_p a_p E_{(p)}$ and furthermore the vielbein coupling (in the bi- and massive gravity cases) becomes equivalent to the metric-language matter coupling proposed by \cite{deRham:2014naa}.

  \item[(b)] {\bf Rank-0 construction}: The `index-carrying' part of the matter coupling is some combination of vielbeins and inverse vielbeins with unit determinant\footnote{This does not mean $\hat M$ needs to be constant, but if any vielbeins are involved in the construction of $\hat M$, the number of type-(p) vielbeins is equal to the number of type-(p) inverse vielbeins, arranged in some (possibly non-trivial) sequence.}, premultiplied by a scalar function of (inverse) vielbeins that encodes spin-2 interactions of the type \eqref{genint}. More specifically this means the matter vielbein takes on the form
\begin{align}
\tilde{E} &= \alpha (E_{(i)}) \hat{M},   &\det \left( \hat{M} \right) &= 1, &\alpha^{D} &= \sum_{i_j}^{\N} \int d^Dx  \; d_{(i_1 i_2 \ldots i_D)} {\cI}_{(i_1 i_2 \ldots i_D)}
\label{eqn:alpha}
\end{align}  
where the constants $d_{(i_1 i_2 \ldots i_D)}$ are independent of the coefficients $c_{(i_1 i_2 \ldots i_D)}$ in the potential ${\cal S}_{\rm pot}$ \eqref{potential}.  In (massive) bi-gravity a specific instance of this case was recently considered in \cite{Heisenberg:2014rka}.
 
  \item[(c)] {\bf Mixed construction}: The effective matter coupling non-trivially depends on the dynamical vielbein \dofs{} through \emph{both} its rank-0 coefficients and its rank-2 constituents, 
\begin{equation}
\tilde{E} =  \frac{\alpha (E_{(i)})}{ \left( \det M \right)^{1/D} }  \, M ( E_{(i)} ) , \;\;\;\; \alpha^D = \sum_{i_j}^{\N} \int d^Dx  \; d_{(i_1 i_2 \ldots i_D)} {\cI}_{(i_1 i_2 \ldots i_D)}
\label{eqn:mixed}
\end{equation}
where $M$ is an arbitrary rank-2 function of the vielbeins. In the massive (bi-)gravity case a coupling of this kind was recently considered in \cite{Heisenberg:2014rka}, in which $\alpha^D$ contained the zeroth order and $D$th order elementary symmetric polynomials only.\footnote{In appendix \ref{appendix:msssumrules} we will see that series expansions of such a coupling would have to satisfy (infinitely many) sum rules (\ref{eqn:msscond3}).}  
  \end{itemize}
All of the above cases yield healthy pure spin-2 interactions of the form \eqref{lambda} by construction (when an effective cosmological constant is present/via matter loops). In the following sections, we assess the ghost-freedom of all these constructions in different limits of the theory -- the results are surmised in Table~\ref{tab:results}.

\section{Constraint analysis}
\label{sec4:constraint}

Having passed the check of generating acceptable pure spin-2 interactions in the section above, a healthy matter coupling should pass a number of other checks. In this section, we will subject our candidate matter couplings to a constraint analysis. If the correct number of constraints is present to allow only the \dofs{} of a (massive) spin-2 field (and the matter fields it couples to) to propagate, then we know that the full theory is free of additional dangerous ghost-like \dofs{} (so-called Boulware-Deser ghosts \cite{Boulware:1973my}). If the theory passes this test and the spin-2 \dofs{} are all healthy (this is related to the Higuchi bound discussion in section \ref{sec6:higuchi}), it is healthy at all energy scales.\footnote{
If it fails, however, we cannot conclude that the theory is sick at all scales. This is because the ghost may sit above the cutoff of the theory, in which case the ghost is simply not present in the low energy regime, where the theory is well-defined. This will be discussed further in section \ref{sec5:decoupling}.} 
Here we will explicitly check for the presence of the correct number of primary constraints -- for a detailed discussion of secondary constraints in related matter couplings see \cite{deRham:2015cha}.

\subsection{ADM decomposition and mini-superspace}


In order to make the Hamiltonian constraint structure of the theory explicit, we would like to express the \dof{} of the theory in terms of canonical coordinates and their momenta. For the vielbeins we consequently follow \cite{Hinterbichler:2012cn} and perform a boosted ADM foliation \cite{Arnowitt:1962hi}, 
\begin{equation}
{E_{(1)}}_{\mu}^{\; A} = \left(
\arraycolsep=1.0pt\def\arraystretch{1.3}
 \begin{array}{c  c}
N\gamma + N^{k}e_{k}^{\;a} P_{a} \;\; & 
\;\; NP^{a} + N^{k} e_{k}^{\; b} \left( \delta_{b}^{a} + \frac{ P_b P^a }{1 + \gamma} \right)   \\
e_{i}^{\; a} P_a & e_{i}^{\;b} \left( \delta_{b}^{a} + \frac{ P_b P^a }{1 + \gamma} \right)
\end{array} \right) , \;\;\;\; {E_{(2)}}_{\mu}^{\; A} = \left( 
\arraycolsep=1.0pt\def\arraystretch{1.3}
\begin{array}{c  c}
M \;\; & \;\; M^{k} l_{k}^{\; a} \\
0 & l_{i}^{\;a} 
\end{array} \right)  ,
\label{eqn:vielbeinadm}
\end{equation}
where the Lorentz $\gamma \equiv \sqrt{1+P_{a}P^{a}}$ and the $D^2$ independent components of a general vielbein ${E_{(1)}}_{\mu}^{\; A}$ are expressed in terms of $(D-1)^2$ components of the spatial vielbein $e_{i}^{\; a}$, the lapse $N$ and $(D-1)$ shifts $N_i$, and the $(D-1)$ momenta $P_a$ (we use the shorthand $P^a_{(1)} \equiv P^a$). Lorentz gauge invariance has been used to fix the momentum for $E_{(2)}$ to $P^{a}_{(2)} =0$.\footnote{In bigravity there is always a single (diagonal) copy of Lorentz invariance remaining, which allows us to do so \cite{Hinterbichler:2012cn}. Since there is only one such copy, it is not in general possible to fix both $P^{a}$ and $P^{a}_{(2)}$, however, so one must leave $P^a$ explicit in this ADM decomposition.}
Forming the corresponding metric from a general vielbein, we find that the dependence on $P_a$ and the $D(D-1)/2$ antisymmetric components of $e_i^{\; a}$ drop out (these are the gauge fields corresponding to Lorentz boosts and rotations). This leaves only the $D (D+1)/2 $ components of a symmetric rank-2 tensor 
\begin{equation}
{E_{(1)}}_{\mu}^{\; A} {E_{(1)}}_{\nu}^{\; A} \eta_{AB} \equiv {g_{(1)}}_{\mu \nu} = \left( 
\arraycolsep=1.0pt\def\arraystretch{1.3}
\begin{array}{c  c}
-N^{2}+N^{k}N_{k} \;\; & \;\;\; N_{j} \\
N_{i} &  \;\;\; g_{ij}
\end{array} \right).
\label{eqn:gadm}
\end{equation}
where $g_{ij} \equiv e_i^{\; a} e_j^{\; b} \delta_{ab}$.
\\

\noindent {\bf General Relativity and \dof{} counting}: In ADM variables, the Einstein-Hilbert kinetic term may be written \cite{Arnowitt:1962hi}, 
\begin{equation}
\sqrt{-g} R [g] = \pi^{ij} \dot{g}_{ij} + N C + N_k C^k + \text{total derivatives},
\label{eqn:EH}
\end{equation}
where the lapse, $N$, and shift, $N^k$, act as Lagrange multipliers, $\pi^{ij}$ is the momentum conjugate to the spatial metric $g_{ij}$ and the precise form of $C$ and $C^k$ is not of importance here. In General Relativity (GR), the usual $D=4$ story is that diffeomorphism invariance can be used to fixed four of the ten independent components of $g_{\mu \nu}$ (effectively this is the statement that the lapse and shift are non-dynamical), leaving six potentially propagating \dofs{} (associated to the spatial metric $g_{ij}$). Another four are rendered non-dynamical by the fact that the lapse and shift in fact enter linearly and generate constraints, $C$ and $C^k$, leaving only two propagating \dofs{} (around a flat background spacetime). Moving from the ghost-free theory of a massless to that of a massive spin-2 field by including the dRGT self-interactions, the lapse and shift still enter non-dynamically leaving six potentially propagating \dofs, but superficially all four shift and lapse constraints are now broken. However, it can be shown that the four $N, N_i$ \eoms{} are not in fact independent \cite{Hassan:2011tf}. This allows a single constraint to remain intact, which eliminates the sixth metric degree of freedom, leaving us with the correct number of \dofs{} for a massive spin-2 field ($2s + 1 = 5$).

When coupling matter directly to the vielbeins, in principle one must verify that enough residual gauge symmetry/constraints exist to ensure none of the $D^2$ components of the general vielbein are ghost-like. In the remainder of this section, we will explicitly consider the lapse and shift constraints. If all $D$ of these are irreparably broken, then (regardless of the Lorentz gauge invariance), we expect the Boulware-Deser ghost \cite{Boulware:1973my} to propagate. 
\\

\noindent {\bf Mini-superspaces}: Before discussing the full ADM analysis, it is instructive to consider the problem in the \emph{mini-superspace}. The mini-superspace describes a consistent subset of solutions of the full ADM decomposition, which are spatially isotropic with $N_i$ and $P_a$ set to zero. Ghost-freedom in this simplified setup represents a check, which any candidate coupling must satisfy in order to have the required number of constraints in the full ADM decomposition. Explicitly, we may write the original vielbeins and the analytic expansion of $\tilde{E}$ as, 
\begin{equation}
\begin{split}
{E_{(1)}}^{(\text{mss})} &= \left( \begin{array}{c c}
N(t,x)     &     0   \\
0      &    a(t,x)
\end{array} \right) ,  \;\;\;\; {E_{(2)}}^{(\text{mss})} = \left( \begin{array}{c c}
M(t,x)     &     0   \\
0      &    b(t,x)
\end{array} \right) , \\
 \implies \tilde{E}^{(\text{mss})} &= \sum_{n,m \in \mathbb{Z}} \alpha_{nm} \left( N, M, a, b \right)  \left( \begin{array}{c c}
N^n M^m     &    0   \\
0      &    a^n b^m 
\end{array} \right) \equiv \left( \begin{array}{c c}
\tilde{N}(t,x)     &     0   \\
0      &    \tilde{a}(t,x)
\end{array} \right) .
\end{split}
\label{eqn:expmss}
\end{equation}
The $\alpha_{nm}$ are fully determined once $\tilde{E}$ is given in terms of $E_{(1)}$ and $E_{(2)}$. 
We emphasise that this implies that the $\alpha_{nm}$ in general result in an explicit lapse dependence for $\tilde{a}$, i.e. all components of the effective matter vielbein in the mini-superspace decomposition then have explicit lapse dependence. As a concrete example, consider, 
\begin{align}
\text{Rank-2:} \;\;\;\; \tilde{E}_\mu^{\ A} &=  E_{(1)}{}_\mu^{\ A} E_{(2)}{}_\nu^{\ B} E_{(1)}^{-1}{}_{\ B}^{\nu}       
&\implies & &\tilde{a} &= b  \label{con1} \\
\text{Rank-0:} \;\;\;\; \tilde{E}_\mu^{\ A} &= \left[ \det \left( E_{(1)} + E_{(2)} \right) \right]^{1/D} \delta_\mu^{\ A}
&\implies & &\tilde{a} &=  \left[ (N+M)(a +b) \right]^{1/D}. \label{con2}
\end{align}
In other words, while $\tilde a$ for the rank-2 construction \eqref{con1} never has any lapse-dependence (as multiplying a string of diagonal matrices cannot mix time-time and space-space entries), it does for the rank-0 construction \eqref{con2} (a non-trivial scalar prefactor must be a fully contracted object, and so contain both $N$ and $a$).

\subsection{Mini-superspaces for Matter Lagrangians}

Depending on the precise matter content, the matter Lagrangian $\mathcal{L}_{M} [\tilde{g}_{\mu \nu}, \Phi_i]$ can affect the constraint structure of the theory in different ways. It is natural to consider a number of known $\mathcal{L}_{M}$ as present in the Standard Model. In this subsection we therefore collect matter Lagrangians (and their corresponding Hamiltonians) for four different kinds of fields, and consider what restrictions each implies on the allowed matter metrics. Note that we will use the shorthand for the effective matter metric $g_{\mu\nu}^{\rm eff} \equiv \tilde g_{\mu\nu}$ throughout. 
\\

\noindent {\bf Vacuum energy}: 
%
First up is the action for an effective cosmological constant, $\tilde{\Lambda}$, \eqref{lambda}, as considered in section \ref{sec3:matterloops} and as will be generated by matter loops, even if there is no bare cosmological constant present in the theory. Trivially this introduces no new (matter) degrees of freedom/conjugate momenta, and so the corresponding Hamiltonian is simply 
\begin{equation}
\mathcal{H}_{\text{mat}} = \det \tilde{E} \; \tilde \Lambda = \tilde{N} \tilde{a} \tilde \Lambda,
\label{mss-lambda}
\end{equation}
where we recall that $\tilde a$ and $\tilde N$ are the scale factor and lapse of the effective matter vielbein respectively.
\\

\noindent {\bf Scalar field}:
%
The action for a scalar field $\chi$, with mass $M$ and minimally coupled to $\tilde{g}$, is
\begin{equation} 
S_{\text{mat}} = \int d^4x\sqrt{-\tilde{g}} \left(  - \frac{1}{2} \tilde{g}^{\mu \nu} \partial_\mu \chi \partial_\nu \chi  - \frac{1}{2}  M^2 \chi^2  \right).
\label{eqn:Lm}
\end{equation}
Accordingly the conjugate momentum for $\chi$ is given by 
\begin{equation}
p_{\chi} \equiv \frac{\delta \mathcal{L}}{\delta \,\partial_0 \chi} =  - \sqrt{-\tilde{g}} \tilde{g}^{\mu 0} \partial_\mu \chi,
\label{eqn:scalarmomentum}
\end{equation}
and the Hamiltonian corresponding to the matter action \eqref{eqn:Lm} consequently reads,
\begin{equation}
\begin{split}
\mathcal{H}_{\text{mat}} = \frac{\sqrt{-\tilde{g}}}{2 \sqrt{\det \tilde{g}_{ij}}} &\Bigg(  \frac{p_\chi^2}{\sqrt{\det \tilde{g}_{ij}}} + \left[  \frac{-\det \tilde{g} }{\sqrt{\det \tilde{g}_{ij}}} \tilde{g}^{k0} \tilde{g}^{l0}  + \sqrt{\det \tilde{g}_{ij}} \tilde{g}^{kl}  \right] \partial_k \chi \partial_l \chi  \\
& \;\;\;\;\;\;\;\;\;\;\;\;\;\;\;\;\;\;\;\;\;\;\;\;\; + \sqrt{\det \tilde{g}_{ij}} M^2 \chi^2   \Bigg) +  \left[ - \frac{\det \tilde{g}}{\det \tilde{g}_{ij} } \tilde{g}^{k0} \right] \partial_k \chi \; p_\chi . \label{eqn:Hm}
\end{split}
\end{equation}
Finally, in the minisuperspace, the effective metric $\tilde{g}$ is diagonal and explicitly substituting in the mini-superspace decomposition \eqref{eqn:expmss} we find
\begin{equation}
\mathcal{H}_{\text{mat}} =  \frac{ \tilde{N} }{ \tilde{a} } \frac{p_\chi^2}{2} + \tilde{N} \tilde{a} \frac{ \left( \partial_k \chi \right)^2}{2} + \tilde{N} \tilde{a} \frac{M^2 \chi^2}{2} . 
\label{eqn:scalarfieldmss}
\end{equation}
\\

\noindent {\bf Yang-Mills field}:
%
Next we consider a matter Yang-Mills field. For simplicity, first consider the Abelian theory of electromagnetism minimally coupled to an effective matter metric $\tilde g$, 
\begin{equation}
S_{\text{mat}} [A ] = - \int d^4 x \sqrt{-\tilde{g}} \left(  \tilde{g}^{\mu \alpha} \tilde{g}^{\nu \beta} F_{\mu \nu} F_{\alpha \beta}  \right).
\end{equation}
The gauge field, $A_\mu$, enters as $F^{\mu \nu} = \nabla^\mu A^\nu - \nabla ^\nu A^\mu$, making the gauge symmetry \\ $A^\mu~\to~A^\mu~+~\nabla^\mu\Lambda$ for a scalar $\Lambda$ explicit. Introducing canonical variables \cite[eqn (II.2.2.2)]{Thiemann:2007zz}, 
\begin{align}
\mathcal{E}^i &= \sqrt{- \tilde{g}} \left(  \nabla_i A_0 + \dot{A}_i \right) , &B^i \left( A \right) &= \epsilon^{ijk} \nabla_j A_k ,
\end{align}
and explicitly substituting a diagonal $\tilde{g}$ via \eqref{eqn:expmss}, gives a classical Hamiltonian, 
\begin{equation}
\begin{split}
\mathcal{H}_{\text{mat}} [ \mathcal{E} , A ] &= \frac{ \sqrt{-\tilde{g}} }{2 \det \left( \tilde{g}_{ij} \right) } \tilde{g}_{ab} \left[ \mathcal{E}^a \mathcal{E}^b  + B^a (A) B^b (A)  \right] , \\
&=  \frac{ \tilde{N} }{ 2 \tilde{a}^2  } \left[ \left( \mathcal{E} \right)^2  + ( B (A) )^{2}  \right] .
\label{MSSYM}
\end{split}
\end{equation}
\par

The extension to a non-Abelian theory is straightforward. The $U(1)$ symmetry is replaced by another group $G$, with multiple generators $T^A$ (one for each dimension of $G$). The above Hamiltonian is modified, 
\begin{align}
A^i &\rightarrow A^i_{\; A} T^A,  &F_{\mu \nu} &\rightarrow F_{\mu \nu}^a = \nabla_{[\mu} A_{\nu]}^a + f^{abc} A_\mu^b A_\nu^c .  
\end{align} 
where $f^{abc}$ are the structure constants of the symmetry algebra, and the gauge fields transform under the adjoint representation of the algebra. However, the mini-superspace lapse dependence remains precisely as in \eqref{MSSYM}. \\

\noindent {\bf Dirac field}:
%
For a fermionic field, $\psi$, of mass $m$, we have a matter action \cite[eqn (2.2)]{Isham:1974ci},  
\begin{equation}
S_{\text{mat}} = \int d^4 x  \sqrt{- \tilde{g}} \left[ \frac{i}{2} \tilde{E}^{\mu}_{\; A} \bar{\psi} \gamma^A \overset{\leftrightarrow}{\nabla}_\mu \psi - m \bar{\psi} \psi \right],
\end{equation}
where we have suppressed the 4-component spinor indices on $\psi$ (they are contracted together via the $4\times 4$ Dirac $\gamma$ matrices). In the mini-superspace, we have,
\begin{equation}
S_{\text{mat}} = \int d^4 x  \det \tilde{E} \left[ \frac{i}{2 \tilde{N}} \left( \bar{\psi} \gamma_0 \dot{\psi} - \dot{\bar{\psi}} \gamma_0 \psi \right) + \frac{i}{2 \tilde{a}} \left( \bar{\psi} \gamma_1 \psi' - \bar{\psi}' \gamma_1 \psi  \right)    -  m  \bar{\psi} \psi    \right]  ,
\end{equation}
where an overdot denotes a time derivative, and a prime denotes a spatial derivative. Using the conjugate momenta, $\pi = \frac{i}{2\tilde{N}} \bar{\psi} \gamma_0  \det \tilde{E}$,  we find a Hamiltonian, 
\begin{equation} 
\mathcal{H}_{\text{mat}} = - \frac{i \tilde{N}}{2} \left(  \bar{\psi} \gamma_1 \psi' - \bar{\psi}' \gamma_1 \psi  \right)      + \tilde{N} \tilde{a} \; m \bar{\psi} \psi.
\label{mss-dirac}
\end{equation}
\\

\ni {\bf Resulting conditions from matter Hamiltonians}: In the minisuperspace picture (\ref{eqn:expmss}), a necessary condition for (Boulware-Deser) ghost freedom is that the lapse constraints be preserved by the coupling. There is no shift to integrate out, and so this amounts to the lapse $N$ appearing linearly in the Hamiltonian as a Lagrange multiplier. Via the matter Hamiltonians \eqref{mss-lambda}, \eqref{eqn:scalarfieldmss}, \eqref{MSSYM} and \eqref{mss-dirac} this puts a constraint on the form taken by $\tilde N$ and $\tilde a$, i.e. the mini-superspace form of the effective matter vielbein. Therefore, to have ghost-free couplings with each of these Standard Model-like matter Hamiltonians requires 
\begin{align}
\text{Vacuum energy:}& \;\;\;\;\;\; \frac{\partial^2}{\partial N^2} \; \tilde{N}  \; \tilde{a}  = 0  , \label{eqn:vaccond} \\
\text{Scalar or Yang-Mills:}& \;\;\;\;\;\; \frac{\partial^2}{\partial N^2} \; \frac{\tilde{N} }{\tilde{a} }  = 0  \;\; \text{and} \;\;  \frac{\partial^2}{\partial N^2} \; \tilde{N} \; \tilde{a}  = 0,   \label{eqn:scalarmss} \\
\text{Dirac spinor field:}& \;\;\;\;\;\; \frac{\partial^2}{\partial N^2} \; \tilde{N}  = 0  \;\; \text{and} \;\;  \frac{\partial^2}{\partial N^2} \; \tilde{N} \; \tilde{a}  = 0 .
\label{eqn:diracmss}
\end{align}
where we recall that $\tilde{N}$ and $\tilde{a}$ are in general functions of $(N, a, M, b)$. Analogous equations hold with derivatives with respect to $M^2$ and $NM$. Only rank-2 constructions (for which $\tilde{a}$ does not depend on $N$) can satisfy all conditions simultaneously -- this can be seen by inspection of equations \eqref{eqn:vaccond}-\eqref{eqn:diracmss}. Rank-0 and mixed constructions (for which $\tilde{a}$ is generically not independent of $N$) cannot be healthy for a general $\mathcal{H}_M$ containing all of the above fields, so cannot be used to couple to matter while respecting the weak equivalence principle.\footnote{Note that, if we were to temporarily ignore the vielbein origins of $\tilde N$ and $\tilde a$ and treat them as independent functions of $N,M,a,b$, one could find a number of solutions satisfying subsets of matter Hamiltonians. For example, in our construction the coupling to vacuum energy is guaranteed, and in addition we could impose the conditions, 
\begin{equation}
\begin{split}
\text{Healthy coupling to Bosons} \;\; &\Leftarrow \;\; \tilde{N} \sim  \sqrt{N}, \; \tilde{a} \sim \sqrt{N}    \\
\text{Healthy coupling to Fermions} \;\; &\Leftarrow \;\; \tilde{N} \sim N ,  \;\;\;\; \tilde{a} \sim \frac{1}{N}  
\end{split}
\end{equation} 
but clearly we may not impose these simultaneously.}
In appendix \ref{appendix:msssumrules} we discuss how \eqref{eqn:vaccond} can also be used to motivate the split into rank-2, rank-0 and mixed matter couplings, which are all `solutions' of \eqref{eqn:vaccond}.

\subsection{Full ADM primary constraints for scalar field matter}

Performing a full ADM decomposition for generic types of matter and general matter couplings is beyond the scope of this paper. However, here we wish to briefly sketch the full ADM decomposition of an effective metric coupling to a massive scalar field (\ref{eqn:Hm}). This simple case already suggests that non-trivial coupling will break some of the constraints in the theory and hence introduce a ghost at some energy scale.

\subsubsection{Rank-2 construction}

By \eqref{eqn:rank2} the single building block rank-2 constructions (i.e. those consisting of a single string from \eqref{bbs} and not of superpositions of those strings) satisfy the following property
\begin{equation}
\det \tilde{E} = \det E_{(p)} =  N_{(p)} \det \left( {e_{(p)}}_{i}^{\; a} \right),
\end{equation}
for some label index $(p)$ by construction. Thus we define $N \equiv N_{(p)}$ and $\sqrt{h} \equiv \det \left( {e_{(p)}}_{i}^{\;a} \right) $ and, in terms of those variables, write the matter Hamiltonian for a massive scalar field in a full ADM decomposition \eqref{eqn:Hm} as  
\begin{equation} \label{eqn:Hscalarmss}
\mathcal{H}_M = \frac{N}{2} \left(  \frac{p_\chi^2}{\sqrt{h}} + \sqrt{h} \left[  N^2 \tilde{g}^{i0} \tilde{g}^{j0}  + \tilde{g}^{ij}  \right] \partial_i \chi \partial_j \chi  + \sqrt{h} M^2 \chi^2   \right) + \left[ N^2 \tilde{g}^{i0} \right] \partial_i \chi\;  p_\chi,
\end{equation}
where $p_\chi$ labels the canonical momentum associated to the scalar field $\chi$ as before. For a non-linear single string coupling as in \eqref{bbs}, we can separate the free spacetime index explicitly, 
\begin{equation}
\tilde{E}_{\mu}^{\; B} = {E_{(s)}}_{\mu}^{\; A} \kappa_A^{\;\, B} , \;\;\;\;
\det \left( \kappa \right) = \begin{cases}
 1 ,\;\;\; & s=p  ,\\
\frac{N_{(p)}}{N_{(s)}} \det\left( e_{(p)} e_{(s)}^{-1} \right), \;\;\; & s \neq p  ,
\end{cases}
\label{eqn:Etildegeneral1}
\end{equation}
where $\kappa$ is constructed from a string of vielbeins. Considering the square brackets of (\ref{eqn:Hscalarmss}), we find\footnote{
Here $E_{(s)}$ is in an upper triangular gauge, $P_{(s)}=0$.
},  
\begin{align}
\frac{N}{2} \sqrt{h} \left[  N^2 \tilde{g}^{i0} \tilde{g}^{j0}  + \tilde{g}^{ij}  \right] &=  \frac{N}{2} \sqrt{h} \Big[  e^i_{\; a} e^j_{\;c} \left( \kappa^a_{\;b} \kappa^{cb} + \kappa^a_{\;B} \kappa^{0B} \kappa^b_{\; C} \kappa^{0C} \right)    \nonumber \\
                                                                                                              &  \;\;\;\;\;\;\;  + \left( 1 + \kappa^0_{\;B} \kappa^{0B} \right) \left( \frac{N^i N^j}{N^2} - \frac{N^{(i} e^{j)}_{\; a} }{N}  \kappa^0_{\; B} \kappa^{a B}     \right)    \Big]  , \label{eqn:sqbrack1}\\
\left[ N^2 \tilde{g}^{i0} \right] &= - N^i \kappa^0_{\; A} \kappa^{0 A} + N e^i_{\; b} \kappa^b_{\; A} \kappa^{0 A}  \label{eqn:sqbrack2}.
\end{align}
\par

The non-linearity in the shift $(1 + \kappa^0_{\; B} \kappa^{0B} ) \partial_i \chi \partial_j \chi N^i N^j $ threatens to re-introduce a Boulware-Deser ghost\footnote{Unless of course $\kappa_A^B = \delta_A^B$ (and we have a healthy minimal coupling) or $\partial_i \chi =0$ (and the matter Lagrangian becomes that of an effective cosmological constant, for which all our couplings are healthy by construction).}. This is because, schematically, the shift equation of motion gives,
\begin{equation}
\frac{\sqrt{h}}{2} (1 + \kappa^0_{\; B} \kappa^{0B} ) \partial_i \chi \partial_j \chi N^i =  N \kappa^0_{\; A}\kappa^{0 A} \partial_i \chi  p_{\chi} + \mathcal{O} \left( \frac{\partial \kappa}{\partial N^j}  \right) \left( \frac{N^i N^j}{N} + ... \right) .
\end{equation}
If $\kappa$ contains some $N_j$ dependence, this becomes a polynomial in $N^i$ with roots $N^i \propto (N)^{t}$, where the power $t$ is not even necessarily integer, let alone unity or zero. If $\kappa$ does not contain any $N_j$ dependence (and is not the identity), then it must mix shifts from the other vielbein species with $N^i$. In both cases, when $N^i$ is integrated out, the resulting Hamiltonian has a non-trivial dependence on the remaining shifts. It is integrating out all these shifts, which then generically introduces a non-linearity in the lapses, eventually leading to the propagation of (an) extra ghost-like degree(s) of freedom.

For the matter couplings of \cite{Noller:2014sta} -- a linear superposition of vielbeins, which in the presence of a symmetric vielbein condition corresponds to the metric construction of \cite{deRham:2014naa} -- \cite{deRham:2014fha,deRham:2015cha} concluded that there is a ghost on integrating out the shift as well.\footnote{
In this case, the Hamiltonian is superficially linear in lapse and shift, however when integrating out the shift, non-linearities in the lapse are induced. This is more subtle than the emergence of a ghost identified for the more complicated \eqref{bbs} couplings, which leads to non-linear shift dependence immediately in the Hamiltonian.} As single non-linear vielbein strings are typically ghostly, and at least linear superpositions also introduce a ghost, this suggests (although of course it does not prove) that \emph{all} non-minimal rank-2 constructions contain a ghost when coupled to a scalar field outside of the mini-superspace. In the following section \ref{sec5:decoupling}, we shall show that this ghost lies above the $\Lambda_3$ decoupling limit scale for the rank-2 constructions and a valid low-energy effective field theory consequently exists, where the mass of the ghost or the energy scale where unitarity breaks down (whichever is lower) will set the cutoff of the theory.

\subsubsection{Rank-0 construction}

Using the rank-0 construction (\ref{eqn:alpha}), we have that (\ref{eqn:Hm}) becomes,
\begin{equation}
\mathcal{H}_M = \frac{1}{2 \alpha^{\frac{D}{2}-1}} \left(  p_\chi^2 + \left[ \alpha^{-1} \hat{M}^{k0} \hat{M}^{l0}  + \alpha^{D-2} \hat{M}^{kl}    \right] \partial_k \chi \partial_l \chi + \alpha^{D-1} M^2 \chi^2   \right)  + \left[ \hat{M}^{k0} \right] \partial_k \chi \; p_\chi. 
\end{equation}
Unless this expression is linear in the lapse, or can be made so by a field redefinition, then we have a propagating potentially ghost-like \dof{} again. 
Outside of the mini-superspace, both $\alpha$ and $\hat{M}$ generally have complicated dependences on $N, N_i$, enforced by the condition (\ref{eqn:alpha}). At least one of the Hamiltonian coefficients will then be irreparably non-linear in lapse and shift,
\begin{equation}
\alpha^D \sim N \;\; \implies \;\;  \frac{1}{\alpha^{D/2}}, \;\; \alpha^{D/2} \;\; \text{non-linear in } N.  
\end{equation}
As these three coefficients multiply different orders of $\chi$ and $p_\chi$, they cannot cancel among themselves. The constraints are broken and the BD ghost is reintroduced. 
The special case of a diagonal $\hat{M}$ warrants attention (for example, if $n_{(p)}=0$ and $\hat{M} = \eta$), as then we have,
\begin{equation}
\mathcal{H}_M = \frac{1}{2 \alpha^{\frac{D}{2}-1}} \left(  p_\chi^2 + \left[ \alpha^{D-2} \right] \partial_k \chi \partial^k \chi + \alpha^{D-1} M \chi^2   \right) . 
\end{equation}
This is manifestly healthy for $M=0, D=2$. In Section~\ref{sec5:decoupling}, we will see that this is consequence of conformal symmetry (a massless scalar field in $D=2$ is conformally invariant \cite{DiFrancesco:1997nk}). However, in the absence of conformal symmetry, all the rank-0 constructions appear to suffer from a ghost when coupled to a scalar field. In the following section, we shall show that this ghost is already present in the $\Lambda_3$ decoupling limit (for rank-0 constructions), and is therefore fatal even in a low-energy \eft{} sense.

\subsubsection{Mixed construction}

The ghost which exists in the scalar field mini-superspace (\ref{eqn:scalarfieldmss}) for a mixed construction coupling (cf. equations \eqref{eqn:vaccond}-\eqref{eqn:diracmss}) cannot be healed by introducing shifts or momenta (leaving the mini-superspace). Therefore without performing the full computation, we can immediately conclude that a mixed construction will have a ghost when coupled to a scalar field.

\section{The $\Lambda_3$ decoupling limit} 
\label{sec5:decoupling}

Given that every non-minimal coupling we have considered so far introduces a ghost in the full ADM decomposition when coupled to a simple massive scalar field, a most pressing question becomes: at what scale does this ghost enter? If the ghost enters at a large enough scale (ideally above the strong coupling scale of the theory), a healthy low-energy \eft{} below that scale is still of interest. If, on the other hand, the ghost is already present at very low scales, one should discard the theory in question. 

In order to answer this question in the context of general matter couplings, we now investigate the $\Lambda_3$ decoupling limit (for the different matter couplings considered so far). For simplicity we will set $M_{(1)} = M_{(2)} \equiv M_{\rm Pl}$ and work in $D=4$ throughout this section, but the conclusions readily generalise. This calculation was carried out for the linear rank-2 linear couplings in the metric formulation by \cite{deRham:2014naa}, and here we extend their approach to the entire set of candidate couplings discussed here.
The $\Lambda_3$ decoupling limit corresponds to performing the following scaling of coupling constants in the theory
\begin{align}
 M_{\rm Pl} &\rightarrow \infty, &m &\rightarrow 0 ,  &\Lambda_3 &\equiv \left(m^2 M_{\rm Pl} \right)^{1/3} \;\; \text{fixed} .
\label{eqn:declimit}
\end{align}
Since we are primarily interested in whether the matter coupling reintroduces a Boulware-Deser like instability, we will focus on the gravitational \dof{} and hence not scale the matter fields.\footnote{Technically this means we keep all further coupling constants that only appear in the matter sector fixed. Note that one may be interested in considering different scaling/decoupling limits to zoom in on interactions at different scales. For example the scale of the high-energy ghost (that lives above the $\Lambda_3$ decoupling limit and is hence not dangerous) in the matter couplings of \cite{deRham:2014naa,Noller:2014sta} could in principle be identified in this way and the same is true for the matter couplings we consider throughout this paper. See \cite{Matas:2015qxa} for a more detailed related discussion. We thank Claudia de Rham and Andrew Tolley for explaining this point to us.}  If interactions in the $\Lambda_3$ decoupling limit are healthy (i.e. ghost-free), then the model in question is a consistent \eft{} (at least) up to the scale $\Lambda_3$. Whether the theory is ghost-free beyond that scale is only a meaningful question if the cutoff of the theory is parametrically larger than $\Lambda_3$.  

In order to make the interactions of different \dofs{} more explicit, we will employ the \St{} trick, that results in an equivalent formulation of the theory with more fields and more symmetry, but the same propagating \dofs{} (the \St{} trick is essentially the inverse of gauge-fixing -- for details in the context of massive, bi- and multi-gravity see \cite{Noller:2013yja,Noller:2015eda}).  

\subsection{\St{} decomposition and interaction scales}

\begin{table}[t]
\begin{center}
\arraycolsep=1cm
\def\arraystretch{1.3}

\begin{tabular}{|c || c | c |}
\hline
Scale & Pure spin-2 interactions & Matter-spin-2 mixing \\
\hline \hline $\Lambda_5$ & $\left( \pi^n \right)$ & $\times$\\ 
$\Lambda_4$ & $ \left( A \pi^{n-1} \right)$ & $\times$\\ 
$\Lambda_3$ & $A^2 \pi^{n-2}, h \pi^{n-1}$ & $\pi^n \Phi_i^m$ \\ 
\vdots & \vdots & $ A \pi^{n-1} \Phi_i^m$ \\ 
\vdots & \vdots & $ A^2 \pi^{n-2} \Phi_i^m$,$ h \pi^{n-1} \Phi_i^m$\\ 
\vdots & \vdots & \vdots \\
\hline
\end{tabular}
\end{center}
\caption{Interaction terms (from the pure spin-2 and matter actions) listed according to the scale they {\it begin} to contribute at, where $\Lambda_\lambda = (M_{\rm Pl} m^{\lambda-1})^{1/\lambda}$. For example, the first row of the table denotes that pure $\pi$ interactions contribute at the scale $\Lambda_5$ and above.  The pure spin-2 interaction is of the form ${\cal S}_{\rm int} = m^2 M_{\rm Pl}^2 \int d^4x E_{(i)} \wedge E_{(j)} \wedge E_{(k)} \wedge E_{(l)}$ while the matter coupling is ${\cal S}_{\rm matter} = \int d^4x (\det \tilde E) {\cal L}[\Phi_i, \tilde g_{\mu\nu}]$. The presence of the dimensionful overall coupling constant $m^2 M_{\rm Pl}^2$ in front of the pure spin-2 interactions and the lack of such a coupling constant in front of the matter action is the reason for the shifted hierarchies. All matter fields $\Phi_i$ minimally couple to the effective metric $\tilde g$ and do not scale with $m$ or $M_{\rm Pl}$. 
%
%
In ghost-free massive (multi/bi)-gravity theories, the interactions of the first two rows vanish up to total derivatives (denoted by brackets in the table), so the least suppressed non-linear interactions that enter from all parts of the action all come in at $\Lambda_3$.}
\label{cubicMG}
\end{table}

Following \cite{Ondo:2013wka,Fasiello:2013woa,deRham:2015cha}, we introduce Lorentz \St{} fields, $\Lambda^A_B$, and diffeomorphism \St{} fields, $Y_\mu$, via the second vielbein\footnote{For a discussion of various equivalent ways of introducing \St{} fields see \cite{Noller:2013yja}. This redundancy is directly related to the existence of Galileon dualities \cite{deRham:2013hsa,deRham:2014lqa,Kampf:2014rka}  and their generalisation to Multi-Galileon dualities \cite{Noller:2015eda}.},
\begin{equation}
{E_{(2)}}_{\mu}^{\; A}  \rightarrow \Lambda^{A}_{\; B} {E_{(2)}}_{\nu}^{\; B} \partial_\mu Y^\nu, 
\label{eqn:Stuckelberging1}
\end{equation}
where
\begin{align}
\Lambda^A_{\ B} &\equiv e^{\lambda^A_{\ B}}, &Y_\mu &\equiv x_\mu - \frac{1}{m M_{\rm Pl}} A_\mu - \frac{1}{\Lambda_3^3} \partial_\mu \pi \label{Apiscale} \\
\lambda_{AB} &\equiv \frac{1}{m M_{\rm Pl}} \hat{\lambda}_{AB}, &\Pi^\nu_{\; \mu} &\equiv \frac{\partial^{\nu}\partial_\mu \pi}{\Lambda_3^3}
\label{eqn:Stuckelberging2}
\end{align}
After performing the \St{} trick, in the decoupling limit $h$, $A$ and $\pi$ capture the dynamics of the helicity-2, -1 and -0 modes respectively. The scaling of $A$ and $\pi$ in \eqref{Apiscale} is set by the coupling constant in front of the graviton potential interactions and their canonical normalisation. The scaling of $\lambda$ is set by the requirement that it enters in the decoupling limit in a non-vanishing and non-divergent way \cite{Ondo:2013wka} -- note that this results in the same scaling for $\lambda$ as for $A$. Performing the \St{} trick also effectively restores the copies of Lorentz and diffeomorphism invariance broken by the massive interaction in the original action. Finally, we will also expand both vielbeins around a (flat) reference vielbein
\begin{align}
{E_{(1)}}_{\mu}^{\; A} &= \delta_{\mu}^{\; A} + \frac{1}{M_{\rm Pl}} {v_{(1)}}_{\mu}^{\; A}, 
&{E_{(2)}}_{\mu}^{\; A} &= \delta_{\nu}^{\; B} + \frac{1}{M_{\rm Pl}} {v_{(2)}}_{\nu}^{\; B}. 
\end{align}

\subsection{Scalar, vector and tensor modes}    

With a potential interaction term like \eqref{potential} the least suppressed non-linear interactions arise at the scale $\Lambda_3$. For our matter couplings to at least give rise to a valid low energy effective field theory with non-trivial dynamics for the graviton, it is therefore important to check that the matter coupling is healthy in the corresponding decoupling limit. In the $\Lambda_3$ decoupling limit the helicity-0, -1, and -2 modes transform as scalar, vector and (rank-2) tensor modes respectively. We will now consider which of these modes contribute to decoupling limit contribution of the matter action and what that contribution is.
\\

\noindent{\bf Hierarchy of modes and decoupling limit vielbeins}: Non-linear interactions of the different helicity modes always enter in the same hierarchy. For a given overall order $n$ in the fields, pure helicity-0 interactions $\pi^n$ enter at the lowest energy scale, followed by $A\pi^{n-1}$ interaction terms which contribute at the next lowest energy scale and which in turn are followed by $h \pi^{n-1}$ and $A^2 \pi^{n-2}$ interaction terms. In particular, this is the case for ghost-free massive (bi-)gravity potential interactions, where the first two sets of terms are tuned to vanish up to total derivatives and the last set of terms comes in at the scale $\Lambda_3$. The $\Lambda_3$ terms then are the least suppressed non-vanishing non-linear interactions in these models. 

When considering a coupling to matter with non-scaling matter fields (as we do here), the hierarchy of scales is still completely set by the scaling of gravitational (helicity) modes. However, two important differences arise. Firstly the graviton potential interaction comes with a $m^2 M_{\rm Pl}^2$ coupling constant, whereas our matter action has no such $m$-dependent coupling constant. This means identical interactions coming from potential vs matter parts of the action can enter at different scales. Secondly, interactions which were tuned to vanish up to total derivatives in the pure spin-2 case, e.g. pure helicity-0 $\pi^n$ interaction terms, can become relevant when mixing with matter \dofs{}. For example,  $\pi^n \Phi_i^m$ terms no longer vanish due to the mixing with matter fields $\Phi_i$. These two observations taken in conjunction mean that, classically and in the absence of an explicit cosmological constant term in the matter action, the non-linear interactions that survive the $\Lambda_3$ decoupling limit are $h \pi^{n-1}$ and $A^2 \pi^{n-2}$ terms from the spin-2 potential part of the action and $\pi^n \Phi_i^m$ from the matter part of the action. No other terms contribute at $\Lambda_3$ and in particular helicity-1 and helicity-2 modes mix with matter fields only at higher energy scales (beyond the decoupling limit). These findings are summarised in table \ref{cubicMG}. Note that, if we were to introduce an overall $m$-dependent coupling constant for the matter action, this would either make the matter contribution to the $\Lambda_3$ decoupling limit vanish or it would introduce non-linear graviton interactions below $\Lambda_3$, in which case one would have to re-do the decoupling limit analysis considered here (non-vanishing non-linear interactions below $\Lambda_3$ diverge in the $\Lambda_3$ limit). Having non-scaling matter fields in the sense discussed above is consequently a straightforward way to have matter fields enter the $\Lambda_3$ decoupling limit in a non-vanishing and non-divergent way.

When computing the matter action contribution to non-linear decoupling limit interactions, we can therefore suppress most of the involved helicity dependence of the vielbeins post-\Sting{} (since this will drop out by the arguments presented above) and instead just work with the `matter decoupling limit' vielbeins
\begin{align} \label{MDLvielbeins}
{E_{(1)}^{\rm MDL}}{}_{\mu}^{\; A} &= \delta_{\mu}^{\; A}, 
&{E_{(2)}^{\rm MDL}}{}_{\mu}^{\; A} &= \delta_{\mu}^{\; A} - \frac{1}{\Lambda_3^3}\partial^A \partial_\mu\pi. 
\end{align}
For an effective matter vielbein $\tilde E = \alpha E_{(1)} + \beta E_{(2)}$, the `matter decoupling limit' of that effective matter vielbein would consequently be $\tilde{E}^{\rm MDL} =  (\alpha + \beta) \mathbb{I} - \beta \Pi$, in matrix notation.
\\

\noindent {\bf Helicity-0 mode \eoms{}}: Seeing as there will be explicit helicity-0 dependence of the decoupling limit matter action, one has to ensure that the dynamics of this mode remains ghost-free in this limit while taking into account the new (matter action) contributions. The equation of motion for the helicity-0 mode $\pi$ in this limit takes on the schematic form 
\begin{equation}
\mathcal{E}_\pi = \frac{\delta S_{\text{pot}}}{\delta \pi} + \frac{\delta S_{\text{mat}}}{\delta \pi} = 0. 
\end{equation}
In the absence of a coupling to matter, the uncoupled dRGT or Bigravity theory yields the ghost-free equations of motion,
\begin{equation}
\mathcal{E}_{\pi} = \frac{\delta S_{\text{pot}}}{\delta \pi} = 0,
\end{equation}
where the $\pi$ \eoms{} are those of a Galileon on flat space \cite{Nicolis:2008in}. Once a coupling to matter is included, the additional contribution from the matter sector may be written, 
\begin{equation}
\begin{split}
\mathcal{E}_\pi^{\mathsf{mat}} &\equiv \frac{\partial S_{\text{mat}} }{\partial \pi} \\[4pt]
&= \frac{1}{\Lambda_3} \partial^\rho \partial_\sigma \left( \frac{\partial S_{\text{mat}}}{\partial \tilde{g}_{\mu \nu}} \frac{\partial \tilde{g}_{\mu \nu}}{\partial \Pi^\rho_{\; \sigma}}  \right) \\[4pt]
&=   \frac{1}{\Lambda_3} \partial^{\rho} \partial_{\; \sigma} \left( \sqrt{-\tilde{g}} \tilde T^{\mu \nu} \frac{ \partial \tilde{E}_{\mu}^{\; C}}{\partial \Pi^{\rho}_{\; \sigma}} \tilde{E}_{\nu}^{\; D} \eta_{CD} \right)  , \;\;\;\; \tilde T^{\mu \nu} \equiv \frac{2}{\sqrt{-\tilde{g}}} \frac{\delta S_{\text{mat}}   }{\delta \tilde{g}_{\mu\nu}}.
\label{eqn:dec1}
\end{split}
\end{equation}
In general, the effective energy-momentum $\tilde{T}_{\mu \nu}$ will depend on derivatives of $\pi$, and one might expect that the $\pi$ \eom{} will consequently depend on higher-than-second-order derivatives of $\pi$, generically introducing an \Ostro{} ghost. While this is true for arbitrary couplings, we shall see that rank-2 and (to some extent also) rank-0 matter couplings investigated here have interesting structure, that allows them to evade the appearance of an (\Ostro) ghost in the $\Lambda_3$ decoupling limit. 

\subsection{Symmetric vielbein condition in the decoupling limit}

We have already argued above that the `matter decoupling vielbeins' (by which we mean the piece of the matter vielbein which survives in the $\Lambda_3$ decoupling limit) take on a vastly simplified form, with the Lorentz \St{} fields, helicity-1 and helicity-2 modes all dropping out. Here we explicitly show this for some specific couplings, demonstrate how this is related to the restoration of the symmetric vielbein condition in the decoupling limit and what this means for superpositions of the vielbein building blocks \eqref{bbs} discussed in this paper.
\\

\ni {\bf Matter vielbeins in the decoupling limit}: To begin, let us consider three specific effective matter vielbeins
\begin{align} \label{specificmatterV}
\tilde E_I{}_{\mu}^{A} &= E_{(1)}{}_{\mu}^{A} + \alpha E_{(2)}{}_{\mu}^{A}, \nn \\
\tilde E_{II}{}^D_{\mu} &= {E_{(1)}}_{\mu}^{\; A} {E_{(2)}}_{\lambda}^{\; B} {E_{(1)}^{-1}}^{\lambda}_{\; C} \eta_{AB} \eta^{DC}, \nn \\
\tilde E_{III}{}^D_{\mu} &= {E_{(2)}}_{\mu}^{\; A} {E_{(1)}}_{\lambda}^{\; B} {E_{(2)}^{-1}}^{\lambda}_{\; C} \eta_{AB} \eta^{DC}, 
\end{align}
We now perform the \St{} replacements \eqref{eqn:Stuckelberging1}, \eqref{eqn:Stuckelberging2} on these matter vielbeins. In order to do so, it is useful to write out the inverse vielbeins (up to quadratic order in the fields) 
\be
\begin{split}
{E^{-1}_{(1)}}^{\mu}_{\, A} &= \delta^\mu_{\, A} - \frac{1}{M_{\rm Pl}} {v_{(1)}}^{\mu}_{\, B} + \frac{1}{2 M_{\rm Pl}^2} {v_{(1)}}^\mu_{\, B} \delta^B_{\, \lambda} {v_{(1)}}^\lambda_{\; A} + {\cal O}\left(v_{(1)}^3\right), 
\end{split}
\ee
\be
\begin{split}
{E^{-1}_{(2)}}^{\mu}_{\, A} &=  \Bigg\{ \delta^\mu_{\; A} + \frac{1}{m M_{\rm Pl}} \partial_A A^\mu + \Pi^\mu_{\; A} \\ 
&\quad+ \frac{1}{M_{\rm Pl}} {v_{(2)}}^\mu_{\; B} \left( - \delta^B_{\; \nu} - \Pi^B_{\; \nu} - \frac{1}{m M_{\rm Pl}} \partial^B A_\nu + \frac{1}{M_{\rm Pl}} {v_{(2)}}^B_{\; \nu}   \right) \delta^\nu_{\; A}  \\
&\quad\quad + \Pi^\mu_{\; B} \Pi^B_{\; \nu} \delta^\nu_{A} + \frac{1}{m M_{\rm Pl}} \partial^B A^{(\mu} \Pi^{\nu)}_{\; B} \delta_{\nu A}     +  \frac{1}{(mM_{\rm Pl})^2} \partial^B A^\mu \partial_B A_\nu \delta^\nu_{\; A} \Bigg\}  \\
&- \frac{1}{m M_{\rm Pl}} \lambda_A^{\; C} \Bigg\{ \delta^\mu_{\; C} + \frac{1}{m M_{\rm Pl}} \partial_C A^\mu + \Pi^\mu_{\; C} + \frac{1}{M_{\rm Pl}} {v_{(2)}}^\mu_{\; C}  \Bigg\} 
+ \frac{1}{2 (m M_{\rm Pl})^2} \lambda_A^{\; B} \lambda_B^{\; \mu} .
\end{split}
\ee
where ${v_{(1)}}^\mu_{\, A} \equiv {v_{(1)}}_{\lambda}^{\, B} \delta^\mu_{\, B} \delta^\lambda_{\, A}$.
As far as the matter decoupling limit is concerned (in fact scaling $M_{\rm Pl} \to \infty$ is sufficient), this means the inverse vielbeins in this limit become
\begin{align} \label{MDL1}
{E_{(1)}^{\rm MDL}}^{-1}{}_{\mu}^{A} &= \delta^\mu_{\; A},
&{E_{(2)}^{\rm MDL}}^{-1}{}^{\mu}_{\, C}    &= \delta^\mu_{\; C} + \Pi^\mu_{\; C} + \Pi^\mu_B \Pi^B_{\; \nu} \delta^\nu_{\; C}. 
\end{align}
Using the explicit examples from \eqref{specificmatterV}, we can also see that performing the decoupling limit scaling in the full effective matter vielbeins has the same effect as substituting using \eqref{MDL1}, as it should. Upon performing the \St{} replacement, the three specific matter vielbeins from \eqref{specificmatterV} become (up to quadratic order in the fields)
\begin{align}
\tilde{E}_I{}_{\mu}^{\; A} &=  (1+\alpha) \delta_\mu^{\, A} + \frac{1}{M_{\rm Pl}} {v_{(1)}}_\mu^{\; A} + \alpha e_2{}_{\mu}^{\; A}     , \\
\tilde{E}_{II}{}_{\mu}^{\; D} &= e_{2}{}_\mu^{\; D} - \frac{1}{2 M_{\rm Pl}^2} {v_{(1)}}_{\mu}^{\; \rho} {v_{(1)}}^{\rho}_{C} \eta^{DC}  \\
\tilde{E}_{III}{}_{\mu}^{\; D} &=  \delta_\mu^{D} + \frac{1}{M_{\rm Pl}}  e_2{}_\mu^{\; A}  {v_{(1)}}_\lambda^{\; B} E_{(2)}^{-1}{}_{\; C}^\lambda \eta_{AB} \eta^{DC}   \\
\nn \text{where} \;\; {e_{2}}_\mu^{\; A} &\equiv 
\Bigg\{ \delta_\mu^{\; A}  - \frac{1}{m M_{\rm Pl}} \partial^A A_{\mu}   - \Pi^A_{\; \mu} + \frac{1}{M_{\rm Pl}} {v_{(2)}}_\mu^{\;A} - \frac{1}{m M_{\rm Pl}^2} {v_{(2)}}_\nu^{\;A} \partial^\nu A_{\mu} - \frac{1}{M_{\rm Pl}} {v_{(2)}}_\nu^{\; A} \Pi^\nu_{\; \mu}   \Bigg\} \\
\nn& + \frac{1}{m M_{\rm Pl}} \lambda^A_{\; B} \Bigg\{ \delta_\mu^{\; B}  - \frac{1}{m M_{\rm Pl}} \partial^B A_{\mu}  - \Pi^B_{\; \mu} + \frac{1}{M_{\rm Pl}} {v_{(2)}}_\mu^{\;B}   \Bigg\} 
+ \frac{1}{2 (m M_{\rm Pl})^2} \lambda^A_{C} \lambda^C_{B} \delta_\mu^{\; B} .
\end{align}
In the $\Lambda_3$ decoupling limit the effective matter vielbeins for the three cases considered therefore become
\begin{align}
\tilde E_I^{\rm MDL}{}_{\mu}^{A} &= (1+\alpha) \delta_\mu^{\, A} - \alpha \Pi_\mu^{\, A} , \nn \\ 
\tilde E_{II}^{\rm MDL}{}^D_{\mu} &= \delta^D_{\mu} - \Pi^D_{\mu}, \nn \\
\tilde E_{III}^{\rm MDL}{}^D_{\mu} &= \delta_\mu^{\; D}.
\label{MDLSpecificvielbeins}
\end{align}
This agrees with the expressions shown in \eqref{MDLvielbeins}, suggesting that the matter decoupling limit version of all the different vielbein building blocks reduce to the matter decoupling limit version of just the linear superposition of 1-vielbein strings, i.e. $\tilde E = \alpha E_1 + \beta E_2$. We will see precisely why this happens when discussing the symmetric vielbein condition below.  
\\

\ni {\bf Restoration of the symmetric vielbein condition}: When inspecting the matter decoupling limit vielbeins in \eqref{MDLSpecificvielbeins}, it is important to notice that the Lorentz \St{} field $\lambda$ completely drops out of the vielbein. Consider what this means for the Lorentz \St{} \eom. This can schematically be written as
\be
\frac{\delta}{\delta \lambda} {\cal S}_{\rm kin} +  \frac{\delta}{\delta \lambda} {\cal S}_{\rm pot} + \frac{\delta}{\delta \lambda} {\cal S}_{\rm mat} = 0.
\ee
Now the kinetic part of the action, the two Einstein-Hilbert terms and their linearised pieces that survive in the decoupling limit, are gauge-invariant, so never contribute to this \eom. What the above scaling arguments show is that, in the decoupling limit, neither does the matter part of the action, so that the decoupling limit version of the $\lambda$ \eom{} is
\be \label{lambdaeomdec}
{\cal E}^{dec}_\lambda \equiv \frac{\delta}{\delta \lambda} {\cal S}^{dec}_{\rm pot} = 0,
\ee
while the general \eom{} is given by
\be
{\cal E}_\lambda \equiv \frac{\delta}{\delta \lambda} {\cal S}^{full}_{\rm pot} + \frac{\delta}{\delta \lambda} {\cal S}^{full}_{\rm mat}= 0.
\label{lambdaeomfull}
\ee
The $\lambda$ \eom{} in the decoupling limit is therefore insensitive to the difference between any of the rank-2 matter couplings (including a minimal coupling) considered throughout this paper. As a result, the well-known result for standard massive (bi-)gravity -- that the symmetric vielbein condition is recovered as a consequence of the $\lambda$ \eom{}\footnote{Note, however, the discussion of other branches of solutions in this context \cite{Deffayet:2012zc,Banados:2013fda}, where the symmetric vielbein condition does not necessarily hold and, in addition, ghost-like instabilities are present.} -- still holds in the decoupling limit version of the theories considered here and equivalence with the metric formulation is obtained as a direct consequence as well. As \eqref{lambdaeomfull}  shows, this will not be the case beyond the decoupling limit.\footnote{Whether this is a physically meaningful region of the theory to explore, depends on the value of the cutoff of the theory.} More specifically the symmetric vielbein (or DvN condition) for two vielbeins takes on the form
\be \label{symvielcond}
\left( E_{(1)}{}^A_\mu E_{(2)}{}^B_\nu -  E_{(1)}{}^A_\nu E_{(2)}{}^B_\mu \right) \eta_{AB} = 0.
\ee
For details on the derivation of this condition from the Lorentz \St{} \eom~and on why this leads to equivalence between the metric and vielbein formulations, we refer to \cite{Ondo:2013wka}.

What is important here is that the commutativity enforced by \eqref{symvielcond} collapses the building blocks of \eqref{bbs} to their simplest form. Consider the following example
\bea
\tilde E_{II}{}^D_{\mu} &=& {E_{(1)}}_{\mu}^{\; A} {E_{(2)}}_{\lambda}^{\; B} {E_{(1)}^{-1}}^{\lambda}_{\; C} \eta_{AB} \eta^{DC} \nn \\ &\to & {E_{(1)}}_{\lambda}^{\; A} {E_{(2)}}_{\mu}^{\; B} {E_{(1)}^{-1}}^{\lambda}_{\; C} \eta_{AB} \eta^{DC} =  {E_{(2)}}_{\mu}^{\; B} \delta^A_C \eta_{AB} \eta^{DC} = {E_{(2)}}_{\mu}^{\; D}
\eea
where we have used the symmetric vielbein condition in going to the second line. Similarly all the building blocks of \eqref{bbs} collapse in the same way when the symmetric vielbein condition is enforced
\begin{align}
\nn \mathsf{1-vielbein} \; &: \;\;  {E_{(p)}}_{\mu}^{\; A} &\to & &{E_{(p)}}_{\mu}^{\; A},\\ 
\nn \mathsf{3-vielbein} \; &: \;\;  {E_{(q)}}_{\mu}^{\; B} {E_{(p)}}_{\lambda}^{\; C} {E_{(q)}^{-1}}^{\lambda}_{\; D} \eta_{BC} \, \eta^{DA}   &\to & &{E_{(p)}}_{\mu}^{\; A}, \\
\mathsf{5-vielbein} \; &: \;\;  {E_{(p)}}_{\mu}^{\;E} {E_{(q)}^{-1}}^{\rho}_{\; E}  {E_{(p)}}_{\rho}^{\; B} {E_{(q)}}_{\lambda}^{\; C} {E_{(p)}^{-1}}^{\lambda}_{\; D} \eta_{BC} \, \eta^{DA} &\to & &{E_{(p)}}_{\mu}^{\; A}. \nn     \\ 
\vdots    \label{bbs-dec}
\end{align}
In other words, all the vielbein building blocks considered collapse to a single vielbein $E_{(p)}$, so coupling to an effective vielbein made up of a single one of any of these building blocks in the decoupling limit reduces to a minimal coupling to $E_{(p)}$. 
\\

\ni {\bf Superpositions}: Earlier we alluded to superpositions of the different vielbein building blocks \eqref{bbs}, but did not consider them in any further detail in sections \ref{sec3:matterloops} and \ref{sec4:constraint}. The considerations in this section now show that such superpositions are in fact also healthy in the decoupling limit. This is because, from \eqref{bbs-dec}, we have that any superposition of the matter vielbein building blocks \eqref{bbs} in the decoupling limit will take on the form of the (matter decoupling limit version of the) linear vielbein superposition of \cite{Noller:2014sta} 
\be \label{MDLsuperposition}
\tilde E_{\rm eff}^{\rm MDL} = \sum_i \alpha_i E_{(i)}^{\rm MDL}, 
\ee
which is known to be healthy at least up to the $\Lambda_3$ scale \cite{deRham:2014naa,deRham:2015cha}. The linear vielbein coupling consequently emerges as the unique form taken by all the rank-2 matter couplings considered in this paper in the decoupling limit, with differences between the different matter couplings only becoming relevant outside of the decoupling limit when the symmetric vielbein condition is no longer enforced (since the $\lambda$ \eom{} now is \eqref{lambdaeomfull} and no longer \eqref{lambdaeomdec}). \par

Superpositions of rank-0 couplings \eqref{eqn:alpha} in the decoupling limit are also particularly simple. Using the commutativity of $E_{(1)}$, $E_{(2)}$, from which $\hat{M}$ is formed, we can annihilate all inverse vielbeins to be left with simply,  
\begin{equation} \label{rank0c}
\tilde{E}^{\text{MDL}}{}_\mu^{\; A} = \alpha \left( E_{(1)}^{\text{MDL}} , E_{(2)}^{\text{MDL}}  \right) \delta_\mu^{\; A},
\end{equation}
where we recall that $\alpha$ is a scalar function of the two vielbeins here and all indices are contracted with $\eta_{\mu}$ and $\eta_{AB}$. Superpositions of such couplings simply redefine $\alpha$.

\subsection{Rank-2 construction}
\label{sec:rank2decoupling}

As discussed above, the symmetric vielbein condition allows us to commute the vielbeins $E_{(1)}$ and $E_{(2)}$, and therefore any single vielbein string satisfying the valence condition \eqref{eqn:rank2} may be reordered so that all inverse vielbeins annihilate, leaving only a single $E_{(1)}$ or $E_{(2)}$ in the case of bigravity. Closely following the metric formulation argument of \cite{deRham:2014naa}, we therefore here show that this coupling is healthy in the vielbein formulation. Allowing for the general linear vielbein superposition \eqref{MDLsuperposition}, we have
\begin{equation}
\begin{split}
\tilde{E}^{\rm MDL}{}_\mu^{\; A} &= c_1 {E_{(1)}^{\rm MDL}}{}_\mu^{\; A} + c_2 {E_{(2)}^{\rm MDL}}_\mu^{\; A} \\
&= \left( (c_1 + c_2) \delta_\mu^{\; A} - c_2 \delta_\nu^{\; A} \Pi^\nu_{\; \mu} \right)  + \mathcal{O} \left( \frac{1}{m M_{\text{Pl}}}  \right) \\
\implies \;\; \mathcal{E}_\pi^{\text{mat}} &\propto \frac{1}{\Lambda_3^3} \partial_C \partial_\mu \left( \sqrt{-\tilde{g}} \tilde{T}^{\mu \nu} \tilde{E}^{\rm MDL}{}_{\nu}^{\; C} \right)
\end{split}
\label{eqn:Emat1}
\end{equation}
Taking the covariant derivative of the stress energy tensor (with respect to the effective metric), we find \cite{Schmidt-May:2014xla, Solomon:2014iwa}, 
\begin{equation}
\begin{split}
 \nabla^{(\tilde{g})}_{\mu} \tilde{T}^{\mu \nu} =  0  \;\;\;\; 
\implies \;\;\;\;  \partial_{\mu} \left( \sqrt{-\tilde{g}} \, \tilde{T}^{\mu \nu}  \right) =  \sqrt{-\tilde{g}} \, \Gamma^\nu_{\alpha \beta} \tilde{T}^{\alpha \beta}
\label{eqn:Tcons}
\end{split}
\end{equation}
where the Christoffel symbols are given by 
$\Gamma^\alpha_{\mu \nu} = - \tilde{g}_{\beta \nu} \partial_\mu \tilde{E}^\beta_{\; C}  \tilde{E}^\alpha_{\; D} \eta^{CD}$ \cite{2006Apei...13..462S}.
In the decoupling limit, this gives, from (\ref{eqn:Emat1}),
\begin{equation}
\begin{split}
\mathcal{E}_\pi^{\text{mat}} &\propto \frac{1}{\Lambda_3^3} \partial_C \left[  \sqrt{-\tilde{g}} T^{\alpha \beta} \Gamma^\nu_{\alpha \beta}   \tilde{E}^{\rm MDL}{}_\nu^{\; C}  + \sqrt{-\tilde{g}} T^{\mu \nu} \partial_\mu \tilde{E}^{\rm MDL}{}_\nu^{\; C}    \right]   \\[4pt]
&= \frac{1}{\Lambda_3^3} \partial_C \left[  - \sqrt{-\tilde{g}} T^{\alpha \beta}  g_{\lambda \beta} \partial_\alpha \tilde{E}^{\rm MDL}{}^{\lambda C}  + \sqrt{-\tilde{g}} T^{\mu \nu} \partial_\mu \tilde{E}^{\rm MDL}{}_\nu^{\; C}    \right] \\[4pt]
&= 0
\end{split}
\end{equation}
All rank-2 couplings, and their superpositions (as discussed above and in section \ref{sec2:cand}), are therefore healthy at and below $\Lambda_3$ when coupled to an arbitrary $\mathcal{L}_{\text{mat}}$ which does not scale with $m$ or $M_{\text{Pl}}$. \\

\noindent {\bf Covariant conservation of $T^{\mu\nu}$}: It is worth emphasising that the covariant conservation of the stress-energy tensor with respect to the effective matter metric \eqref{eqn:Tcons} follows as a consequence of the weak equivalence principle-respecting matter coupling, and does not rely on any further details of the matter coupling. Take the following matter Lagrangian for explicitness
\be
{\cal L}_{\rm mat} = -\frac{1}{2}\sqrt{-g_{\rm eff}}\left(g_{\rm eff}^{\mu\nu}\pa_\mu \chi \pa_\nu \chi + M^2 \chi^2 \right)
\ee
The corresponding matter \eom, i.e. in this case simply the \eom{} for $\chi$, is given by \cite{deRham:2014naa}
\be \label{weak}
\epsilon_\chi = \Box_{\rm eff} \chi - M^2 \chi = \nabla^{\rm eff}_\mu T^{\mu\nu}= 0.
\ee
The key observation is that the only way in which $\chi$ knows about the gravity sector is via the minimal coupling to $g_{\rm eff}$. So regardless of what form $g_{\rm eff}$ takes, whether it is St\"uckelberged or not, whether it is replaced by a different effective matter metric or not, \eqref{weak} will remain valid with respect to the effective metric that matter is minimally coupled to. We have explicitly shown this for a single scalar field here, but none of the reasoning depends on this assumption.

\subsection{Rank-0 construction}

Suppose the effective matter vielbein coupling takes a rank-0 form \eqref{eqn:alpha}. In the decoupling limit, the effective vielbein becomes \eqref{rank0c}, and we now have 
\begin{equation}
\mathcal{E}_\pi^{\text{mat}} = \frac{1}{\Lambda_3^3} \partial^\rho \partial_\sigma \left( \sqrt{-\tilde{g}} \, 
\tilde{T}^\mu_{\; \mu} \, \frac{\partial \alpha}{\partial \Pi^{\rho}_{\; \sigma}} \alpha  \right) .
\label{eqn:rank0decouplingcond}
\end{equation}
This no longer vanishes in general on-shell.  However, in special cases where the energy-momentum is traceless, \eqref{eqn:rank0decouplingcond} trivially vanishes and the coupling \eqref{rank0c} consequently does not introduce a ghost. Note that systems with traceless energy-momentum are scale invariant, and (together with a virial condition \cite{DiFrancesco:1997nk}) this implies conformal symmetry. 

While this means the rank-0 coupling is disqualified as a universal coupling for all matter, let us briefly consider a coupling to a system with conformal symmetry. As a concrete example consider a Klein-Gordon scalar field $\phi$ minimally coupled to $\tilde{g}$. This corresponds to the following matter Lagrangian,
\begin{equation} \label{eqn:scalarfieldT}
\begin{split}
\mathcal{L}_{\text{mat}} [\phi , \tilde{g}] &=  \tilde{g}^{\mu \nu} \partial_\mu \bar{\phi} \partial_\nu \phi - m^2 \bar{\phi} \phi  \\[4pt]
\implies \;\; T^{\mu \nu} &= \left( \tilde{g}^{\mu \alpha} \tilde{g}^{\nu \beta}  +  \tilde{g}^{\mu \beta} \tilde{g}^{\nu \alpha}  -  \tilde{g}^{\mu \nu} \tilde{g}^{\alpha \beta}  \right) \partial_\alpha \bar{\phi} \partial_\beta \phi - \tilde{g}^{\mu \nu} m^2 \bar{\phi} \phi  \\[4pt]
\implies \;\; T^\mu_\mu &=  \left( 2 - D \right) \tilde{g}^{\alpha \beta} \partial_\alpha  \bar{\phi} \partial_\alpha \phi  -  D m^2 \bar{\phi} \phi  .
\end{split}
\end{equation}
This shows that general scalar fields will not have vanishing $T^{\mu \nu}$ trace, and so a scalar field coupled to gravity via the rank-0 coupling as discussed here will generically give rise to a ghost, even in the $\Lambda_3$ decoupling limit. For a Klein-Gordon scalar the exception is now readily seen to be a massless scalar field around flat (1+1)-dimensional space---a well-known example of conformal symmetry.
However, coupling individual matter species (in some given dimension) to gravity in this way, whilst others have to be coupled in some different way, of course does away with the weak equivalence principle altogether. Such an approach was already partially discussed in \cite{Matas:2015qxa}, but we shall leave fully exploring the consequences of opening this Pandora's box for future work. 

For general matter, without a traceless energy-momentum tensor, the non-vanishing contribution to the $\pi$ \eoms{} will generically be higher-derivative, introducing an \Ostro{} ghost. This is the case because $\partial \alpha/\pa \Pi$ generically depends on $\Pi$ and \eqref{eqn:rank0decouplingcond} consequently depends on up to fourth derivatives acting on $\pi$, leaving us with the conclusion that  rank-0 matter couplings are generically ghostly in the decoupling limit, even if their `effective cosmological constant'/one-(matter)loop contributions are ghost-free (see section \ref{sec3:matterloops}).

\subsection{Mixed construction}

Mixed constructions do not have any helpful index structure, and typically (for general $\alpha$ and superpositions) have an \Ostro{} ghost in the helicity-0 components of the metrics below $\Lambda_3$, 
\begin{equation}
\mathcal{E}_{\pi}^{\rm mat}  =  \frac{1}{\Lambda_3} \partial^{\rho} \partial_{\; \sigma} \left( \sqrt{-\tilde{g}} T^{\mu \nu} \frac{ \partial \tilde{E}^{\rm MDL}{}_{\mu}^{\; C}}{\partial \Pi^{\rho}_{\; \sigma}} \tilde{E}^{\rm MDL}{}_{\nu}^{\; D} \eta_{CD} \right) . 
\end{equation}
The set of matter Lagrangians to which the mixed constructions can healthily couple is given by a restricted set of stress-energy tensors, namely those for which,
\begin{equation*}
\partial^{\rho} \partial_{\; \sigma} \left( \sqrt{-\tilde{g}} T^{\mu \nu} \frac{ \partial \tilde{E}_{\mu}^{\; C}}{\partial \Pi^{\rho}_{\; \sigma}} \tilde{E}_{\nu}^{\; D} \eta_{CD} \right) 
\end{equation*}
is at most second order in field derivatives in the decoupling limit. This is more restrictive than either the rank-2 set (all $\mathcal{L}_M$) or the rank-0 set (conformal $\mathcal{L}_M$), and employing such a coupling would require doing away with the weak equivalence principle even more radically than in the rank-0 case. 

The exceptional cases are where the mixed $\tilde{E}$ actually reduces to a rank-2 coupling in the decoupling limit, when the symmetric vielbein condition can be imposed. This occurs whenever,
\begin{equation}
\alpha ( E_{(i)}^{\text{MDL}} ) = ( \det M (E_{(i)}^{\rm MDL} ) )^{1/D} .
\end{equation}
As a concrete example, consider, 
\begin{equation}
\begin{split}
\tilde{E} &= \left( \frac{\det \left( E_{(1)} + E_{(2)} \right) }{ \det \left( E_{(1)} + E_{(1)} E_{(2)} E_{(1)}^{-1}  \right) } \right)^{1/D}  \, \left( E_{(1)} + E_{(1)} E_{(2)} E_{(1)}^{-1} \right) \\
\implies \;\; \tilde{E}^{\rm MDL} &= E_{(1)} + E_{(2)}  
\end{split}
\end{equation}
Indeed, any pair of rank-2 constructions from Section \ref{sec3:matterloops} can be used in place of $E_{(1)}, E_{(1)} E_{(2)} E_{(1)}^{-1}$ and this kind of construction (and any constant superposition of such terms) will be healthy in the decoupling limit, as per the arguments of Section \ref{sec:rank2decoupling}.  

\section{The Higuchi bound} 
\label{sec6:higuchi}

Until this point we have performed various checks for the presence of a Boulware-Deser ghost in massive (bi-)gravity models with a non-trivial matter coupling. In this section we will consider the Higuchi ghost instead, i.e. we will check that the theories in question are well-defined in the sense that perturbations of the helicity-0 mode $\pi$ are well-defined around cosmologically motivated FRW backgrounds. This was checked for minimally coupled bigravity in \cite{Fasiello:2013woa} and we extend this analysis to the case of the general matter couplings considered in this paper.
The Higuchi bound, representing the condition for the absence of a Higuchi ghost, can be derived in various ways, namely via a full Hamiltonian analysis, via the mini-superspace or via a decoupling limit analysis\footnote{For a discussion of all of these approaches within the contexts of massive gravity see \cite{Fasiello:2013woa}.}. Here we closely follow the mini-superspace approach of \cite{Fasiello:2013woa} in order to check  whether one may simultaneously satisfy the Higuchi bound \cite{Higuchi:1986py} and have a working Vainshtein mechanism \cite{Vainshtein:1972sx, Babichev:2013usa} and what conditions this imposes. 

In general there is a tension between these two requirements. This is because the Higuchi bound relates the value of the Hubble parameter $H$ to the graviton mass $m$, while simultaneously the observed proximity to a GR-like solution (the Vainshtein mechanism allows this to take place in massive (bi-)gravity theories) on large scales means $H$-dependent corrections to the Friedmann equations should be small for large $H$. Generically the resulting bounds can be incompatible -- indeed this is the case for FLRW solutions in massive gravity \cite{Fasiello:2012rw} -- so it is important to verify whether the bigravity theories discussed here can simultaneously be free of Higuchi ghosts and lead to acceptable cosmological (background) evolutions.

In this section we will consider two vielbeins, $E_{(g)}$ and $E_{(f)}$, in the $D=4$ dimensional FLRW ansatz, 
\begin{equation}
\begin{split}
ds_g^2 &= -N^2 dt^2 + a(t)^2 d\mathbf{x}^2 , \\
ds_f^2 &= -M^2 dt^2 + b(t)^2 d\mathbf{x}^2.
\end{split}
\label{eqn:FRLW}
\end{equation}
The corresponding Hubble rates associated with each metric are then, 
\begin{equation}
H_g^2 \equiv \frac{\dot{a}^2}{N^2 a^2} , \;\;\;\; H_f^2 \equiv \frac{\dot{b}^2}{M^2 b^2} .
\end{equation}
This ansatz allows us to use the mini-superspace vielbeins \eqref{eqn:expmss} as discussed in Section \ref{sec4:constraint}. We recall that there we found that general rank-2 couplings are potentially healthy, rank-0 couplings may only be healthy if the matter has traceless stress-energy, and mixed constructions are generally not healthy (so will not be considered in this section). 

Schematically, one can consider (on introducing the \St{} fields as in Section~{\ref{sec5:decoupling}) expanding to second order in helicity-0 perturbations about FLRW,  
\begin{equation}
\delta^2 S  \supset  F(m, H_g, H_f)  \delta \dot{\pi}^2
\end{equation} 
where $F$ is a function which depends on the choice of theory. Requiring that these perturbations in $\pi$ are not ghost-like immediately gives the Higuchi bound,
\begin{equation}
F(m, H_g, H_f) \geq 0 .
\end{equation}
However, by restricting attention to a certain region of $(H_g , H_f)$ parameter space, it may be possible to remove the $m$ dependence from the Higuchi bound---removing the Higuchi-Vainshtein tension in the process. This is certainly the case for massive bigravity with a simple single-metric coupling to matter \cite{Fasiello:2013woa}. 
In order to have valid FLRW solutions, both this bound and a complementary condition on $m$ from observed cosmic expansion history must be satisfied. The latter requires a continuity with GR, which is only possible in massive gravity with a Vainshtein mechanism. \cite{Fasiello:2012rw} discusses a tension between these two requirements (Higuchi bound and Vainshtein regime) for massive gravity in detail, concluding that there are no stable and empirically viable FLRW solutions (for massive gravity). 

We will study how the form of $F$, and the allowed region of $(H_g, H_f)$ parameter space, varies when our different matter couplings are used as part of the action,
\begin{equation} \label{H1}
\begin{split}
S &= V_3 \int dt \left\{ \; \frac{M_g^2}{2} R[E_{(g)}] + \frac{M_f^2}{2} R[E_{(f)}] - m^2 M_g^2 \det E_{(g)} \sum_{n=1}^3 \beta_n e_n \left( E_{(g)}^{-1} E_{(f)}  \right)   \right\} + S_{\text{mat}} \\
S_{\text{mat}} &= - V_3 \int dt \, \det \tilde{E} \, \mathcal{L}_{\text{mat}} [\tilde{g}, \phi] 
\end{split}
\end{equation}
where $V_3$ is a constant spatial volume, $e_n$ are the usual characteristic symmetric polynomials. For generic $\beta$ parameters interesting distinct regimes where the bound is satisfied can be found. Exploring the full Higuchi parameter-space is beyond the scope of this paper (we do so in \cite{Higuchi}). But to illustrate that several of the couplings discussed throughout this paper can be healthy, we here focus on models with $\beta_0 = 0 = \beta_4$ (hence the sum in \eqref{H1} only runs $1-3$ in other words, we set the cosmological constants for each metric/vielbein to zero). We emphasise that \eqref{H1} this is simply rewriting \eqref{S-gen} in order to enable an easier comparison with the derivation of \cite{Fasiello:2013woa}. In FLRW variables \eqref{eqn:FRLW}, this action becomes, 
\begin{equation}
S = V_3 \int dt \left\{  N a^3 \left(  -3 M_g^2 \frac{\dot{a}^2}{N^2 a^2} - \mathcal{B} \right) + M b^3 \left( - 3 M_f^2 \frac{\dot{b}^2}{ M^2 b^2} - \frac{a^3}{b^3} \mathcal{A}  \right) \right\} + S_{\text{mat}},
\label{eqn:higuchiSselfint}
\end{equation}
where the dRGT $\beta_n$ coefficients enter as,  
\begin{equation}
\begin{split}
\mathcal{B} &= m^2 M_g^2 \left( 3 \beta_1 \frac{b}{a} + 3 \beta_2 \left( \frac{b}{a} \right)^2 + \beta_3 \left(  \frac{b}{a} \right)^3 \right) \\
\mathcal{A} &= m^2 M_g^2 \left(  \beta_1 + 3 \beta_2 \left( \frac{b}{a} \right) + 3 \beta_3 \left(  \frac{b}{a} \right)^2 \right) . 
\end{split}
\end{equation}
For different $D$, these are simply modified. Note that we will define the dimensionless field, $e^\chi \equiv b/a$, such that $\delta \chi \sim \delta \dot{\pi}$. For details on how $\chi$ is related to the \St{} procedure and why it captures kinetic interactions of (perturbations of) the helicity-0 $\pi$ in this way, we refer to \cite{Fasiello:2013woa}.
In what follows we will now explicitly consider a matter Lagrangian for a free scalar field, 
\begin{align} \label{eqn:fiducialmatter}
\mathcal{L}_{\text{mat}} &=  \frac{1}{2} \tilde{g}^{\mu \nu} \partial_\mu \phi \partial_\nu \phi 
&\to & &\mathcal{L}_{\text{mat}} &= \frac{1}{2} \tilde{g}^{00} \dot{\phi}^2  \;\; \text{if homogeneous,} 
\end{align}
and note when our results are independent of this choice. 

\subsection{Minimal matter couplings: A warm-up}

We first briefly recap the derivation of the Higuchi bound for a minimal coupling of matter to a single metric/vielbein. We closely follow the derivation and notation of \cite{Fasiello:2013woa}. For a minimal coupling $\tilde{E} = E_{(g)}$, 
\begin{equation}
S_{\text{mat}} = V_3 \int dt \, a^3 \frac{\dot{\phi}^2}{2 N}  .
\end{equation}
We can then replace time derivatives with the canonically conjugate momenta (going to a Hamiltonian formulation), 
\begin{equation}
p_a = -6 M_g^2 \frac{\dot{a}}{N} a , \;\;\;\;
p_b = -6 M_f^2 \frac{\dot{b}}{M} b , \;\;\;\;
p_\phi = a^3 \frac{\dot{\phi}}{N}
\end{equation}
\begin{equation}
S = V_3 \int dt \left\{ 
p_a \dot{a} + p_b \dot{b} + p_\phi \dot{\phi} 
- N \left( - \frac{p_a^2}{12 M_g^2 a} + a^3 \mathcal{B} + \frac{p_\phi^2}{2 a^3} \right)
- M \left( - \frac{p_b^2}{12 M_f^2 b} + a^3 \mathcal{A}  \right)     
\right\} . 
\end{equation}
The lapse constraints then correspond to the Friedmann equations, 
\begin{equation}
3 M_g^2 H_g^2 = \mathcal{B} + P^2 , \;\;\;\; 3 M_f^2 H_f^2 = e^{-3 \chi}  \mathcal{A}  , \;\;\;\; P^2 = \frac{p_\phi^2}{2 a^6}
\label{eqn:Hparams}
\end{equation}
and can be used to integrate out $p_a$ and $p_b$. The equation of motion for $p_\phi$ is then, 
\begin{equation}
p_\phi \dot{\phi} = \sqrt{12} a^2 M_g \frac{ P^2 }{ \sqrt{B + P^2} } 
\end{equation}
which can be used to integrate $p_{\phi}$ out also. This leaves, in the convenient gauge $a \propto t$,  
\begin{equation}
\begin{split}
S &= 6 t^2 M_g^2 V_3 \int dt \left\{ - \sqrt{\frac{B}{3 M_g^2}} \sqrt{1 - \frac{t^2 \dot{\phi}^2}{6 M_g^2} } -   \frac{1+\dot{\chi}t}{\sqrt{3 M_g^2}} \frac{M_f}{M_g}  \sqrt{e^{3\chi} \mathcal{A}}    \right\}   \\
&\equiv V_3 \int dt \, \mathcal{L} [ \chi, \dot{\chi}, g ] , \;\;\;\; g^2 \equiv \frac{t^2 \dot{\phi}^2}{6 M_g^2} 
\end{split}
\end{equation}
where $g$ is a dimensionless field proportional to the matter energy density (not to be confused with any property of $g_{\mu \nu}$). \par

We can now perturb the action to second order, 
\begin{equation}
\delta^2 S = V_3 \int dt \sum_{q_1, q_2 \in \left\{\chi, g\right\}} \left\{ \frac{\partial^2 \mathcal{L}}{\partial q_1 \partial q_2} \delta q_1 \delta q_2 + 2 \frac{\partial^2 \mathcal{L}}{\partial q_1 \partial \dot{q}_2} \delta q_1 \delta \dot{q}_2 + \frac{\partial^2 \mathcal{L}}{\partial \dot{q}_1 \partial \dot{q}_2} \delta \dot{q}_1 \delta \dot{q}_2   \right\} .
\label{eqn:d2S}
\end{equation}
Note that $\partial \mathcal{L} / \partial \dot{g} = \partial^2 \mathcal{L} / \partial \dot{\chi}^2 = 0$, and that the $2 (\partial^2 \mathcal{L} / \partial \chi \partial \dot{\chi} ) \delta \chi \delta \dot{\chi}$ term can be integrated by parts, giving, 
\begin{equation}
\delta^2 S = V_3 \int dt \left\{  
\left( \frac{\partial^2 \mathcal{L}}{\partial \chi^2} - \frac{d}{dt} \frac{\partial^2 \mathcal{L}}{\partial \chi \partial \dot{\chi}} \right) \delta \chi^2 
+ 2 \frac{\partial^2 \mathcal{L}}{\partial \chi \partial g} \delta \chi \delta g 
+ \frac{\partial^2 \mathcal{L}}{\partial g^2}  \delta g^2  \right\} .
\end{equation}
This can be diagonalised by the linear, invertible field redefinition,
\begin{equation}
\delta g = \delta \bar{g} - \frac{ \partial^2 \mathcal{L} / \partial \chi \partial g }{\partial^2 \mathcal{L} / \partial g^2} \delta \chi , 
\label{eqn:diagonalisation}
\end{equation}
giving the desired Higuchi bound, 
\begin{equation}
\begin{split}
\delta^2 S &\supset V_3 \int dt \, F \delta \chi^2  + \mathcal{O} (\delta g) \\
 F &=  \frac{\partial^2 \mathcal{L}}{\partial \chi^2} - \frac{d}{dt} \frac{\partial^2 \mathcal{L}}{\partial \chi \partial \dot{\chi}}  - \frac{ \left( \partial^2 \mathcal{L} / \partial \chi \partial g \right)^2 }{\partial^2 \mathcal{L} / \partial g^2}  \geq 0 . \\
\end{split}
\end{equation}
as we require the sign of the kinetic fluctuations for $\delta \pi$ to be positive.

In order to write $F$ solely in terms of $(m, H_g, H_f)$ we need to solve for $\chi (m, H_g, H_f)$. The field equation of motion is, 
\begin{equation}
\frac{2}{ \det E_{(g)} } \frac{\delta S}{\delta g^{\mu \nu}} = M_g^2 G_{\mu \nu} [E_{(g)}] + m^2 M_{g}^2 \sum_{n=0}^3 (-1)^n \beta_{n} Y_{\mu \nu}^{(n)} - T_{\mu \nu} 
\end{equation}
where the $Y_{\mu \nu}^{(n)}$ are given in \cite{Schmidt-May:2014xla}. On taking the covariant derivative with respect to $E_{(g)}$ and imposing stress-energy conservation and the Bianchi identity, this gives \cite{Enander:2014xga}, 
\begin{equation}
\nabla_{(g)}^{\mu} \frac{1}{\sqrt{-g}} \frac{\delta S}{\delta g^{\mu \nu}} \propto m^2 M_g^2 \left( \beta_1 a^2 + \beta_2 a b + \beta_3 b^3   \right) \left[ M \dot{a} - N \dot{b} \right] . 
\label{eqn:minimalBianchi}
\end{equation}
This has two branches of solution. Setting the first (rounded) bracket to zero gives a cosmology which is indistinguishable to GR at all scales at the background level \cite{vonStrauss:2011mq}, but is infinitely strongly coupled at the level of perturbations \cite{Gumrukcuoglu:2011zh, Comelli:2012db,D'Amico:2012pi}. 
We will focus on the second branch\footnote{
This is called the normal branch in \cite{Fasiello:2013woa}, or branch 2 in \cite{Comelli:2012db}.
}, setting the second (square) bracket to zero. This gives the desired relation, 
\begin{equation}
e^\chi = \frac{H_g}{H_f} .
\label{eqn:minimalfieldeom}
\end{equation}

Using \eqref{eqn:minimalfieldeom} to write $F$ solely in terms of the Hubble rates, we have 
(up to an unimportant overall constant factor of $6 t^2 M_g^2$)
, 
\begin{equation}
F(m, H_g, H_f) = \frac{\tilde{m}^2}{4 H_g} \left( 
\frac{\tilde{m}^2 H_f^2}{ H_g^4} \frac{M_g^2}{M_f^2} + \frac{\tilde{m}^2}{ H_g^2 } - 2
\right)  \geq 0 
\end{equation}
where $\tilde{m}$ is the so-called `dressed mass', 
\begin{equation}
\tilde{m}^2 = m^2   \left( \beta_1  \frac{H_g}{H_f} + 2 \beta_2 \frac{H_g^2}{H_f^2} + \beta_3 \frac{H_g^3}{H_f^3}   \right) . 
\end{equation}
Note that, as expected, this is in complete agreement with the result\footnote{Specifically this agrees with equation (3.18) of \cite{Fasiello:2013woa}, with $H_g \equiv H$. 
} of \cite{Fasiello:2013woa}. 

Now, tension with the Vainshtein mechanism is resolved by setting $H_f \gg H_g$, so that $e^{\chi}$ is a perturbatively small parameter. To leading order, the Friedman equations \eqref{eqn:Hparams} then imply,
\begin{equation}
1- g^2  \sim \frac{\beta_1 m^2}{ H_f H_g} , \;\;\;\; H_f \sim H_g \frac{3 H_g^2}{\beta_1 m^2} \frac{M_f^2}{M_p^2}
\label{eqn:Hleadingorder}
\end{equation}
with required consistency conditions, $\frac{\beta_1 M_p^2}{3 M_f^2} \ll H_g$, $M_f > 0$. These conditions are violated by the massive gravity limit ($M_f \to 0$, so $E_{f}$ non-dynamical), indicating that this special $H_f \gg H_g$ regime of parameter space only exists in the bigravity case. The Higuchi bound in this regime is,
\begin{equation}
\frac{\beta_1 m^2}{4 H_f} \left\{ 1 + \frac{H_g}{H_f} \frac{ \beta_1 m^2}{H_g^2} + \mathcal{O} \left(  \frac{H_g}{H_f}   \right)    \right\}  \geq 0 
\label{eqn:Higuchiboundminimal}
\end{equation}
and is satisfied to leading order for all $m^2 > 0$, and therefore consistent with a Vainshtein mechanism and $m \to 0$ GR limit. We have included the next-to-leading-order $m^2/H_g H_f$ for comparison with the general rank-2 case, which will have a larger next-to-leading-order term (of $H_g^2/m^2 < 1$).  

Note that the dependence on the actual form of $\mathcal{L}_{\text{mat}}$ drops out of the final inequality. Therefore, for minimal coupling to arbitrary matter, there is no Higuchi ghost (helicity-0 instability about FLRW) at any value of $m^2$, including in the GR limit $m \to 0$.

\subsection{General rank-2 couplings}

With the FLRW ansatz \eqref{eqn:FRLW}, we can use the mini-superspace vielbeins \eqref{eqn:expmss}. These are diagonal, and therefore commute---and so all rank-2 constructions collapse to, 
\begin{equation}
\tilde{E} = c_1 E_{(g)} + c_2 E_{(f)}.
\end{equation}
Further, we can absorb one of these constant coefficients into the definition of the fields in $\mathcal{L}_{\text{mat}}$, and consider the coupling $\tilde{E} = E_{(1)}+ c_2 E_{(2)}$. This gives a matter action,
\begin{equation}
S_M = V_3 \int dt \, \left( a + c_2 b \right)^3 \frac{ \dot{\phi}^2 }{ 2 (N + c_2 M) }  .
\end{equation}
As before, we go to a Hamiltonian formulation in order to integrate out the momenta,
\begin{equation}
p_a = -6 M_g^2 \frac{\dot{a}}{N} a , \;\;\;\;
p_b = -6 M_f^2 \frac{\dot{b}}{M} b , \;\;\;\;
p_\phi = (a+c_2 b)^3 \frac{\dot{\phi}}{N+c_2 M}
\end{equation}
\begin{equation}
\begin{split}
S = V_3 \int dt \Bigg\{ 
p_a \dot{a} + p_b \dot{b} + p_\phi \dot{\phi} 
&- N \left( - \frac{p_a^2}{12 M_g^2 a} + a^3 \mathcal{B} + \frac{p_\phi^2}{2 (a+c_2 b)^3} \right) \\
&- M \left( - \frac{p_b^2}{12 M_f^2 b} + a^3 \mathcal{A} +  \frac{c_2 p_\phi^2}{2 (a+c_2 b)^3}  \right)     
\Bigg\} . 
\end{split}
\end{equation}
The lapse constraints give Friedmann equations, 
\begin{equation}
3 M_g^2 H_g^2 = \mathcal{B} + P^2 , \;\;\;\; 3 M_f^2 H_f^2 = e^{-3 \chi} \left( \mathcal{A} + c_2 P^2 \right) , \;\;\;\; P^2 \equiv \frac{p_\phi^2}{2 (a +c_2 b)^6}
\label{eqn:Hparams2}
\end{equation}
and can be used to integrate out $p_a$ and $p_b$. The equation of motion for $p_\phi$ is then, in the gauge $a \propto t$, 
\begin{equation}
\begin{split}
p_\phi \dot{\phi} &= \sqrt{12} t^2 M_g \frac{ P^2 }{ \sqrt{\mathcal{B} + P^2 } } \left( 1 + \epsilon  \right) \\
\epsilon &\equiv (1 + \dot{\chi} t ) \frac{c_2 M_f}{M_p} \sqrt{\frac{\mathcal{B} +P^2 }{e^{-3 \chi} ( \mathcal{A} + c_2 P^2 )   }}  = (1+\dot{\chi}t ) \frac{c_2 H_g}{H_f}
\end{split}
\label{eqn:epsilonP}
\end{equation}
which can be used to integrate $p_{\phi}$ out also. This leaves,
\begin{equation}
\begin{split}
S &= 6 t^2 M_g^2 V_3 \int dt \left\{ - \sqrt{\frac{B}{3 M_g^2}} \sqrt{1 - \frac{g^2}{1+\epsilon^2} } -   \frac{1+\dot{\chi} t}{\sqrt{3 M_g^2}} \frac{M_f}{M_g} e^{\frac{3}{2}\chi} \mathcal{A} \sqrt{ \frac{1 - \frac{g^2}{1+\epsilon^2}}{A (1 - \frac{g^2}{1+\epsilon^2}) + c_2 B \frac{g^2}{1+\epsilon^2}  } }   \right\}   \\
&\equiv V_3 \int dt \, \mathcal{L}_{\epsilon} [ \chi, \dot{\chi}, g ] , \;\;\;\; g^2 \equiv \frac{t^2 \dot{\phi}^2}{6 M_g^2 (1 + c_2 e^{\chi})^3} 
\end{split}
\label{eqn:Lepsilon}
\end{equation}
where $g$ again is a dimensionless field proportional to the matter energy density (not to be confused with any property $g_{\mu \nu}$). 

Before we perturb the final version of the action (with $p_\phi$ integrated out), let us consider the field equations, 
\begin{equation}
\frac{2}{ \det E_{(g)} } \frac{\delta S}{\delta g^{\mu \nu}} = M_g^2 G_{\mu \nu} [E_{(g)}] + m^2 M_{g}^2 \sum_{n=0}^3 (-1)^n \beta_{n} Y_{\mu \nu}^{(n)} - \frac{\det \tilde{E}}{\det E_{(g)}} T^{\alpha \beta} \frac{ \delta \tilde{g}_{\alpha \beta}}{\delta g^{\mu \nu}} 
\end{equation}
where the $Y_{\mu \nu}^{(n)}$ are given in \cite{Schmidt-May:2014xla}. On taking the covariant derivative with respect to $E_{(g)}$ and using stress-energy conservation and the Bianchi identity, this gives \cite{Enander:2014xga}, 
\begin{equation}
\nabla_{(g)}^{\mu} \frac{1}{\sqrt{-g}} \frac{\delta S}{\delta g^{\mu \nu}} \propto \left( m^2 M_g^2 (\beta_1 a^2 + \beta_2 a b + \beta_3 b^3) - c_2 a^2 p   \right) \left[ M \dot{a} - N \dot{b} \right] . 
\end{equation}
where $p$ is the pressure exerted by the matter fields. Note that the stress energy is covariantly conserved with respect to $\tilde{g}_{\mu \nu}$, not $g_{\mu \nu}$, which gives the new pressure dependent term. Again, this has two branches of solution---in this case setting the first (rounded) bracket to zero can give interesting cosmologies \cite{Solomon:2014iwa, Comelli:2015pua, Gumrukcuoglu:2015nua, Lagos:2015sya}, however we will continue to focus on the second branch, \eqref{eqn:minimalfieldeom}, as this has a well-defined $c_2 \to 0$ limit.

Following the example of a minimal coupling, we specialise to the regime $H_f \gg H_g$, and ask whether this remains a region of parameter space in which there is no Higuchi ghost (and allows a working Vainshtein mechanism). Once more, this means that $e^\chi$ is a perturbatively small parameter, and so the Friedman equations \eqref{eqn:Hparams2} to leading order,
\begin{align} \label{eqn:doublycoupledscaling}
\frac{3 M_f^2}{M_g^2} \frac{H_g}{H_f} \frac{H_g^2}{m^2} - \frac{3 c_2 g^2}{(1+\epsilon)^2} \frac{H_g^2}{m^2} &\sim \beta_1 + ... , \\
3 \left( 1- \frac{g^2}{(1+\epsilon^2)^2} \right) \frac{H_f}{H_g} \frac{H_g^2}{m^2} &\sim 3 \beta_1 + ... ,
\end{align}
can give the same leading order behaviour \eqref{eqn:Hleadingorder}, \emph{providing} we have the consistency conditions $\frac{\beta_1 M_g^2}{3M_f^2} \ll \frac{H_g^2}{m^2} \ll \frac{\beta_1}{3 c_2 g^2}$. Once again the lower bound forbids the $M_f \to 0$ massive gravity limit, but now there is also an upper limit, which disappears in the limit $c_2 \to 0$. If this upper bound is exceeded, then $H_f$ acquires a leading order dependence in $1/c_2$, and hence there is no smooth $c_2 \to 0$ limit. 

Further, depending on the fixed value of $c_2$, it is possible that we have a second perturbatively small parameter, $H_g^2 / m^2$. In this work we shall allow $c_2$ to be general, and make the choice\footnote{
The other ordering, $\frac{H_g}{H_f} \gg \frac{H_g^2}{m^2}$, required for $\frac{\beta_1}{c_2 g^2} \ll 1$, may produce qualitatively very distinct behaviour, and this is a region of $(H_g, H_f, c_2)$ parameter space worth exploring in future work. 
} that $H_g / H_f$ remains the smallest value in the problem,   
\begin{equation}
\frac{H_g}{H_f} \ll \frac{H_g^2}{m^2}
\end{equation}  
This has the implications, 
\begin{equation}
g^2 \sim 1 , \;\;\;\;  1  \ll \frac{H_f}{H_g} \ll \frac{3 M_f^2}{| \beta_1 | M_g^2} \;\; \implies \;\; M_f > \sqrt{\frac{|\beta_1|}{3}} M_p .
\end{equation}
In this self-consistent $H_f \gg H_g$ regime, we can expand the Lagrangian in powers of $\epsilon \propto H_g/H_f$, 
\begin{equation}
\begin{split}
\mathcal{L} &= \sum_{n=0}^{\infty} \frac{\partial^n \mathcal{L}_{\epsilon} }{\partial \epsilon^n}\Big|_{\epsilon=0} \frac{\epsilon^n}{n!} \\
\frac{\partial^2 \mathcal{L}_{\epsilon}}{\partial q_1 \partial q_2} &= \sum_{n=0}^{\infty} \frac{1}{n!} \frac{\partial^n}{\partial \epsilon^n}\Big|_{\epsilon=0} \Bigg( 
 \frac{\partial^2 \mathcal{L}_{\epsilon}}{\partial q_1 \partial q_2} \epsilon^n 
+ n \frac{\partial \mathcal{L}_{\epsilon}}{\partial q_1} \epsilon^{n-1} \frac{\partial \epsilon}{\partial q_2}
+ n \frac{\partial \mathcal{L}_{\epsilon}}{\partial q_2} \epsilon^{n-1} \frac{\partial \epsilon}{\partial q_1} \\
& \quad\quad\quad\quad\quad\quad\quad\quad + 
\mathcal{L}_{\epsilon} \left( n (n-1) \epsilon^{n-2} \frac{\partial \epsilon}{\partial q_1} \frac{\partial \epsilon}{\partial q_2}   + n \epsilon^n \frac{\partial^2 \epsilon}{\partial q_1 \partial q_2} \right)
\Bigg)
\end{split}
\label{eqn:epsilonexp}
\end{equation}
By considering the relative order of each of these terms,
as well as the diagonalisation \eqref{eqn:diagonalisation}, we find that to leading order\footnote{
Note that the $\chi$ dependence of $g$ \eqref{eqn:Lepsilon} is subleading order,
\begin{equation}
\delta g = \delta g|_{\chi=0} + \frac{\partial g}{\partial \chi} \delta \chi = \delta g|_{\chi=0} + \mathcal{O}(\frac{H_g}{H_f}) ,
\end{equation}
and so the second order expansion of the action \eqref{eqn:d2S} is still valid.
},
\begin{equation}
F =  \frac{\partial^2 \mathcal{L}_{0}}{\partial \chi^2} - \frac{d}{dt} \frac{\partial^2 \mathcal{L}_{0}}{\partial \chi \partial \dot{\chi}}  - \frac{ \left( \partial^2 \mathcal{L}_{0} / \partial \chi \partial g \right)^2 }{\partial^2 \mathcal{L}_{0} / \partial g^2}  \geq 0 ,
\end{equation}
where, 
\begin{equation}
\mathcal{L}_{0} \propto - \sqrt{ \frac{B}{3 M_g^2} } \sqrt{1 - g^2} - \frac{(1+ \dot{\chi} t)}{\sqrt{3 M_g^2}} \frac{M_f}{M_g} \sqrt{e^{3 \chi}} \left( 1 + \frac{c_2 g^2}{1 - g^2} \frac{B}{A} \right)^{-1/2} .
\end{equation}
This gives, to leading order, the same Higuchi bound as the minimally coupled case \eqref{eqn:Higuchiboundminimal},
\begin{equation} \label{ourHiguchi}
\frac{\beta_1 m^2}{4 H_f } \left\{ 1 + \frac{H_g^2}{m^2} \frac{4 c_2}{\beta_1} \left( \frac{15}{8} + \dot{\chi} t  \right) + \mathcal{O} \left( \frac{H_g}{H_f} \frac{m^2}{H_g^2}  \right)   \right\} \geq 0 .
\end{equation} 
However, at next-to-leading order, this is a new result, with an explicit matter dependence entering via $\dot \chi$. Note that the next-to-leading-order term vanishes in the $c_2 \to 0$ limit, and so has not appeared in previous calculations of this bound for singly coupled matter. While this term (and the other parameter space regions we have not considered) may have interesting physical consequences, we will postpone a detailed discussion to future work. Significantly, in the context of our new matter couplings, we have shown that there exists a region of parameter space ($H_f \gg H_g, \; M_f > \sqrt{|\beta_1|/3} M_p$) in which our new rank-2 couplings may be expanded about FLRW without an Higuchi ghost interfering with a Vainshtein mechanism and a smooth GR $m \to 0$ limit. 

\begin{figure}
\centering
\includegraphics[width=0.75\textwidth]{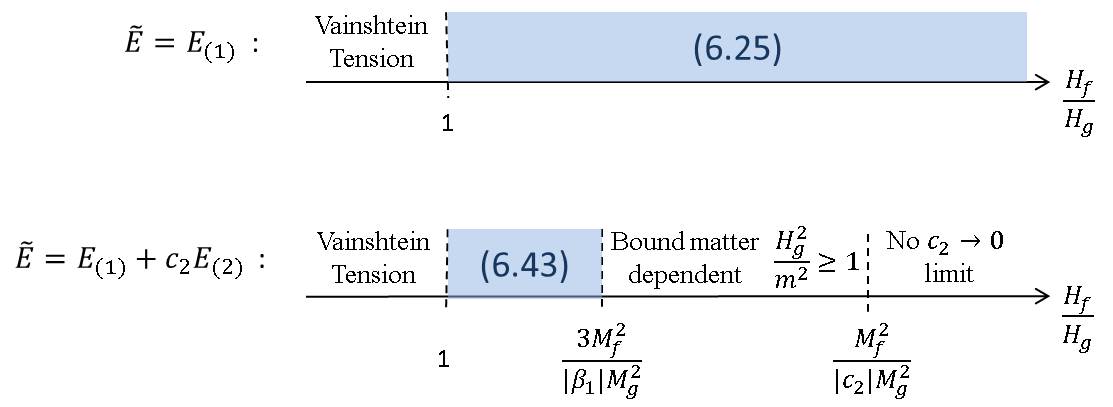}
\caption{Regimes of $(H_g,H_f)$ parameter space where our perturbative calculation holds are shaded. Other regions may also admit stable FLRW solutions or interesting phenomenology, but we content ourselves with demonstrating that, for $M_f >\sqrt{|\beta_1|/3 \;}\, M_g$, there are at least some viable solutions. \eqref{eqn:doublycoupledscaling} shows that, for certain choices of $H_f/H_g$, a matter ($g$) dependence or a singular $c_2\to0$ limit is introduced.}
\end{figure}

\subsection{Rank-0 constructions}

Any rank-0 coupling in the FLRW ansatz may be written as, 
\begin{equation}
\tilde{E}_\mu^{\; A} = \alpha \delta_\mu^{\;A} , \;\;\;\; \alpha^D = \det E_g \sum_{n=0}^4 \gamma_n e_n ( E_{(g)}^{-1} E_{(f)} ) 
\end{equation}
The field equations of motion are,
\begin{equation}
\begin{split}
\frac{2}{\det E_{(g)}} \frac{\delta S}{\delta g^{\mu \nu}} = M_g^2 G_{\mu \nu} [E_{(g)}] &+ m^2 M_g^2 \sum_{n=0}^3 (-1)^n \beta_n Y_{\mu \nu}^{(n)}  \\
& - \frac{\det \tilde{E}}{\det E_{(g)}} T^{\alpha \beta} \left( \sum_{n=0}^4 \gamma_n e_n ( E_{(g)}^{-1} E_{(f)} )  \right)  \left( \sum_{n=0}^3 (-1)^n \gamma_n Y_{\mu \nu}^{(n)}  \right) \eta_{\alpha \beta}
\end{split}
\end{equation}
(and analogously for the $E_{(f)}$ variation). Note that the term proportional to the stress-energy no longer preserves the $M \dot{a} = N \dot{b}$ branch studied in the rank-2 case, \emph{unless} the stress-energy is traceless, in which case its contribution vanishes, and we are left with the same Bianchi constraint \eqref{eqn:minimalBianchi} as in the minimal coupling case. We will consider such a traceless energy-momentum, and focus once more on the branch \eqref{eqn:minimalfieldeom}. 

Using the free scalar field \eqref{eqn:fiducialmatter} requires $D=2$ in order to have conformal symmetry. In this case the Higuchi bound is always trivially satisfied because a massive spin-2 field carries no degrees of freedom, the theory is purely topological.  Staying in $D=4$ we instead consider the Maxwell action for a time-dependent vector field $A_{\mu} (t)$ in temporal gauge $A_0=0$, 
\begin{equation}
\begin{split}
S_{\text{mat}} &= \int d^4 x \, \det \tilde{E} \; \tilde{g}^{\mu \rho} \tilde{g}^{\nu \sigma}  \left(- \frac{1}{4}  F_{\mu \nu} F_{\rho \sigma} \right)  = V_3 \int dt \, \alpha^D  \; \frac{\eta^{\mu \rho}}{\alpha^2} \frac{\eta^{\nu \sigma}}{\alpha^2}  \frac{1}{2} \left( \partial_\mu A_\nu \partial_{[\rho} A_{\sigma ]} \right)   \\
&= V_3 \int dt \, \frac{1}{2} \dot{A}_{i} \dot{A}_{i}  ,
\end{split}
\end{equation}
and so once again we can change to canonical momenta,
\begin{equation}
p_a = -6 M_g^2 \frac{\dot{a}}{N} a , \;\;\;\;
p_b = -6 M_f^2 \frac{\dot{b}}{M} b , \;\;\;\;
p_{A_i} = \dot{A}_i
\end{equation}
\begin{equation}
\begin{split}
S = V_3 \int dt \Bigg\{ 
p_a \dot{a} + p_b \dot{b} + p_{A_i} \dot{A}_i
&- N \left( - \frac{p_a^2}{12 M_g^2 a} + a^3 \mathcal{B}  \right) \\
&- M \left( - \frac{p_b^2}{12 M_f^2 b} + a^3 \mathcal{A}   \right)     -  \frac{p_{A_i}^2}{2}
\Bigg\} . 
\end{split}
\end{equation}
Note that the matter sector has completely decoupled from the metric fields, and will therefore evolve independently. The calculation of the Higuchi bound then goes through exactly as before, this time with the trivial condition $P^2 =0$. 

As such the rank-0 couplings (with $\beta_0 = 0$) may produce stable FLRW solutions in the specific case where this coupling is only to external matter with traceless energy-momentum (as would be implied by conformal symmetry). However, the ghost identified in the mini-superspace will enter in this kind of FRW setup whenever $T_{\mu \nu}$ is not traceless - and so, as discussed above, use of this coupling amounts to a violation of the strong equivalence principle.


\section{New couplings and uniqueness}
\label{sec7:unique}

Arguments for the uniqueness of the matter couplings of \cite{deRham:2014naa,Noller:2014sta} were recently presented in \cite{deRham:2015cha,Huang:2015yga,Heisenberg:2015iqa,Matas:2015qxa}. So how do the new couplings presented here fit into these arguments? Two aspects of the uniqueness arguments are important to highlight in this context and in order to explain how our new couplings escape these arguments:
\\

\ni {\bf Metric vs vielbein formulation}: In \cite{Huang:2015yga,Heisenberg:2015iqa} the matter couplings of \cite{deRham:2014naa,Noller:2014sta} were argued to be unique in the metric formulation, in the sense that no other matter couplings remain ghost-free at least up to the scale $\Lambda_3$. All the new consistent couplings we have discussed in this paper (i.e. the rank-2 constructions) reduce back to the couplings of \cite{deRham:2014naa,Noller:2014sta} whenever the symmetric vielbein condition is imposed, as we have shown above. In other words, when the symmetric vielbein condition holds and consequently an equivalent metric formulation exists for our couplings, our results are perfectly compatible with the claims of \cite{Huang:2015yga,Heisenberg:2015iqa}. Only when an equivalent metric formulation ceases to exist, i.e. in settings where the analyses of \cite{Huang:2015yga,Heisenberg:2015iqa} do not apply, do our new couplings become different. 
\\

\ni {\bf Jordan frame}: The linear vielbein coupling has the interesting feature that the associated field re-definition which maps the action from Einstein to Jordan frame -- a linear field re-definition of the vielbeins --  also leaves the form of the potential-like bigravity interactions invariant \cite{Noller:2014ioa,Schmidt-May:2014xla}. In other words, this field re-definition changes the values of coupling constants in the non-derivative bigravity interactions, but not their overall form. If (a superposition of) the vielbein building blocks in \eqref{bbs} is used, this will of course no longer be the case and the Jordan frame transformation will no longer leave the potential interactions invariant. This was pointed out in \cite{deRham:2015cha,Matas:2015qxa}. It is important to keep in mind that, in order to determine where in the energy spectrum, if anywhere, any potential ghosts are located, one now has to take into account the nature of the modified potential as well as of the now modified kinetic terms (the Jordan frame transformation is not a symmetry of the Einstein-Hilbert terms). The non-standard form of the potential interactions in the Jordan frame is not enough to conclude that another coupling is ghostly (at least up to the scale $\Lambda_3$). This is the case if one works in the Jordan frame -- when working in the Einstein frame kinetic and potential interactions of course take on their usual form here.
In any case, as we have explicitly shown above, the new rank-2 couplings introduced here only lead to differences to the couplings of \cite{deRham:2014naa,Noller:2014sta} outside of the decoupling limit, so that the decoupling limit proof of the healthiness of the couplings of \cite{deRham:2014naa,Noller:2014sta} in \cite{deRham:2014naa} also covers the couplings considered here (given that the metric and vielbein formulations are identical in this limit as discussed in sections \ref{sec2:cand} and \ref{sec5:decoupling}).

\section{Conclusions} 
\label{sec:conc}

In this paper we considered different ways of coupling matter to spin-2 fields. Until recently the minimal GR-like coupling of a metric/spin-2 field was the only known consistent way to do so without introducing extra \dofs{} by hand, i.e. the only known consistent way of universally coupling a spin-2 field to all other fields in the theory. The recently proposed couplings of \cite{deRham:2014naa,Noller:2014sta} add to the space of consistent couplings in a low-energy \eft{} sense - they are consistent couplings at least up to the scale $\Lambda_3$. 

Here we have systematically considered a variety of further couplings constructed order-by-order in the vielbein formulation. In particular we have identified a family of new consistent universal matter couplings where matter minimally couples to a metric 
\be
\tilde g^{\rm eff}_{\mu\nu} \equiv \tilde E_{\rm eff}{}^A_\mu \tilde E_{\rm eff}{}^B_\nu \eta_{AB},
\ee
where the effective matter vielbein $\tilde E$ is an arbitrary superposition of any of the following terms
\begin{align}
\nn \mathsf{1-vielbein} \; &: \;\;  {E_{(p)}}_{\mu}^{\; A},\\ 
\nn \mathsf{3-vielbein} \; &: \;\;  {E_{(q)}}_{\mu}^{\; B} {E_{(p)}}_{\lambda}^{\; C} {E_{(q)}^{-1}}^{\lambda}_{\; D} \eta_{BC} \, \eta^{DA}, \\
\mathsf{5-vielbein} \; &: \;\;  {E_{(p)}}_{\mu}^{\;E} {E_{(q)}^{-1}}^{\rho}_{\; E}  {E_{(p)}}_{\rho}^{\; B} {E_{(q)}}_{\lambda}^{\; C} {E_{(p)}^{-1}}^{\lambda}_{\; D} \eta_{BC} \, \eta^{DA}, \nn     \\ 
\vdots  
\end{align}
and so on. Here $(p),(q)$ are simply label indices and the individual terms are all built from the two vielbeins $E_{(1)}$ and $E_{(2)}$ in the theory.
As such, in index-free notation we can write the effective matter-coupling vielbein as
\be \label{vielsup}
\tilde E_{\rm eff} = \sum_i \alpha^{(1)}_i E_{(i)} + \sum_{i,j}^{i \neq j} \alpha^{(3)}_{i,j} {E_{(j)}} {E_{(i)}} E_{(j)}^{-1} + \sum_{i,j}^{i \neq j} \alpha^{(5)}_{i,j} {E_{(i)}} E_{(j)}^{-1} {E_{(i)}}  {E_{(j)}} E_{(i)}^{-1} + \ldots
\ee
This extends the known consistent matter couplings of \cite{deRham:2014naa,Noller:2014sta}, which took the form $\tilde E_{\rm eff} = \sum_i \alpha_i E_{(i)}$ and hence were the lowest order piece of \eqref{vielsup}. Matter couplings of this type, as presented in this paper, have the following key features
\begin{itemize}
\item They are healthy at least up to the energy scale $\Lambda_3$, as shown by the $\Lambda_3$ decoupling limit analysis in section \ref{sec5:decoupling}, and are therefore part of a {\it consistent low-energy \eft}. In this limit, the symmetric vielbein is also restored and as a result matter couplings of the type \eqref{vielsup} effectively reduce to the consistent couplings of \cite{deRham:2014naa,Noller:2014sta} {\it in this limit}. 
\item In the {\it massive gravity limit}, where only one vielbein is dynamical, \eqref{vielsup} automatically collapses back to the construction of \cite{deRham:2014naa,Noller:2014sta}, also beyond any decoupling/scaling limit. Extensions of our matter coupling construction to {\it Multi-Gravity theories} with more spin-2 fields are straightforward and discussed explicitly in appendix \ref{appendix:multi}.
\item The effective cosmological constant generated by a matter coupling of this type, and consequently matter loop corrections to this coupling, generates healthy dRGT-type pure spin-2 interactions, whenever the effective vielbein is made up from just a single of the building blocks in \eqref{bbs}. Equivalently, this is the case when only one of the coefficients in \eqref{vielsup} is non-zero. For general superpositions like \eqref{vielsup}, this will not be the case. However, by the decoupling limit argument above, this should only significantly affect the theory above $\Lambda_3$. 
\item The Higuchi bound for these couplings can be satisfied in a certain region of parameter space (for a more complete analysis see \cite{Higuchi}) and in the space considered in this paper the bound agrees with that derived for minimal couplings \cite{Fasiello:2013woa} to leading order. At next-to-leading order we do find modifications \eqref{ourHiguchi}, however.
\item A minisuperspace analysis for these couplings shows them to be healthy, also indicating that healthy FLRW solutions exist at the background level. A full constraint analysis was not carried out here, but, just as for the linear vielbein coupling of \cite{deRham:2014naa,Noller:2014sta}, we expect there to be a ghost-like instability at some scale $\Lambda_{\rm ghost} > \Lambda_3$, effectively imposing a cutoff for the theory.
\end{itemize}
In addition to the couplings \eqref{vielsup}, which we dubbed `rank-2' couplings, we also briefly investigated what we called `rank-0' and `mixed' couplings, essentially effective vielbein couplings where there is an overall scalar function of the gravitational \dofs{} multiplying some rank-2 object. For details we refer to sections \ref{sec3:matterloops}--\ref{sec6:higuchi}, but to summarise we found that matter cannot be coupled to these constructions universally without introducing a ghost already at the $\Lambda_3$ scale and in the minisuperspace, \emph{unless} the couplings reduce back to the rank-2 constructions in this decoupling limit (in the mixed case), or if the coupling is restricted to particular types of matter with conformal symmetry (in the rank-0 case).

These findings raise several interesting questions for future research. First and foremost the cutoff scale $\Lambda_c$ for this whole class of spin-2 field theories coupled to matter has to be established along the lines of \cite{Matas:2015qxa} in order to know how far beyond $\Lambda_3$ they represent a valid low-energy \eft. If $\Lambda_{\rm cut} \gg \Lambda_3$, it will be of great interest to investigate the phenomenological differences between the different couplings \eqref{vielsup} presented here. In general this is also true of differences between the vielbein and metric formulation of massive (bi)-gravity theories with non-trivial matter coupling, as discussed throughout this paper and in \cite{deRham:2015cha,Hinterbichler:2015yaa}. Finally it would be interesting to prove whether the couplings presented here really are the unique set of couplings in the vielbein language that yield consistent field theories up to at least $\Lambda_3$, or if further extensions exist (and if so, where they differ). For example, it may be possible to find additional matter coupling terms that simply vanish in the decoupling limit and only contribute above it. Answering some (or all) of these questions should move us closer to fully understanding the space of consistent spin-2 field theories, and in the process improve our understanding of the way gravity works.
\\
 
\noindent {\bf Acknowledgements:} We would like to thank James Scargill, Claudia de Rham, Pedro Ferreira, Macarena Lagos and Andrew Tolley for comments on drafts of this paper and several helpful discussions, Lavinia Heisenberg and Raquel Ribeiro for comments on a draft and James Bonifacio and Angnis Schmidt-May for interesting related discussions. SM is supported by the von Clemm Fellowship. JN is supported by the Royal Commission for the Exhibition of 1851, BIPAC and Queen's College Oxford.

\pagebreak
\appendix

\let\oldaddtocontents\addtocontents \renewcommand{\addtocontents}[2]{}

\section{Extension to multi-gravity theories} 
\label{appendix:multi}

Given that we have been working in the vielbein formulation throughout, a generalisation of our matter couplings to setups with $N$ spin-2 fields is straightforward. For potential interactions the full consistent multi-gravity generalisation of the ghost-free potential massive and bigravity interactions was presented in \cite{Hinterbichler:2012cn} -- features of multi-gravity theories are further discussed in \cite{Khosravi:2011zi,Hinterbichler:2012cn,Hassan:2012wt,Nomura:2012xr,Tamanini:2013xia,Noller:2013yja,Scargill:2014wya,Goon:2014paa,Noller:2015eda}.

In the multi-gravity case the effective matter vielbein that matter can couple to is made up of the following building blocks
\begin{align}
\nn \mathsf{1-vielbein} \; &: \;\;  {E_{(p)}}_{\mu}^{\; A},\\ 
\nn \mathsf{3-vielbein} \; &: \;\;  {E_{(q)}}_{\mu}^{\; B} {E_{(p)}}_{\lambda}^{\; C} {E_{(q)}^{-1}}^{\lambda}_{\; D} \eta_{BC} \, \eta^{DA}, \\
\mathsf{5-vielbein} \; &: \;\;  {E_{(r)}}_{\mu}^{\;E} {E_{(q)}^{-1}}^{\rho}_{\; E}  {E_{(p)}}_{\rho}^{\; B} {E_{(q)}}_{\lambda}^{\; C} {E_{(r)}^{-1}}^{\lambda}_{\; D} \eta_{BC} \, \eta^{DA}, \nn     \\ 
\vdots  
\label{bbs-multi}
\end{align}  
just like in \eqref{bbs}, except that, at higher than cubic order in the vielbeins and their inverses, we can now have more than two species of vielbeins. These are labelled by $p,q,r,\ldots$ and, except for $p$ in the above expressions have to satisfy $n_q-\bar n_{q} = 0, n_r-\bar n_{r} = 0$ and so on, where $n_i$ labels the number of vielbeins $E_{(i)}$ in the building block and $\bar n_i$ labels the number of inverse vielbeins $E_{(i)}^{-1}$ in the building block. 
We impose this restriction, because with it the mini-superspace action will be linear in the lapse associated with $(p)$, $N_{(p)}$. Otherwise we will have a dependence of the mini-superspace action on $N_{(p)}^{n} M_{(p)}^{m} O_{(p)}^{o} \ldots $ with the restriction that all the powers $n+m+p+\ldots = 1$, which is imposed by the index structure of the effective vielbein (essentially by requiring one free Lorentz and one free space-time index to survive). If more than one power i non-zero, the mini-superspace will already be highly non-linear in the lapses and the ghost-free constraint structure of the theory will generically be destroyed.

The nature of the decoupling limit and symmetric vielbein condition is somewhat more complicated in multi-gravity. The decoupling limit was investigated in \cite{Noller:2013yja,Noller:2015eda} and the symmetric vielbein condition is known to be broken in the presence of `cycles of interactions' \cite{Scargill:2014wya,deRham:2015cha}, leading to an inequivalence between the metric and vielbein formulation even at the level of the decoupling limit. In fact, in the metric formulation, they even lower the scale of the least suppressed non-linear interactions \cite{Scargill:2014wya}. In the absence of such cycles of interactions -- equivalently: loops in the theory graph -- we expect the symmetric vielbein condition to be recovered. When building a tree graph of interactions for a multi-gravity theory with only bigravity type interactions, this is straightforward to see \cite{Hinterbichler:2012cn} and each interaction gives the relevant symmetric vielbein condition. Proving this statement rigorously in the general case, i.e. that the symmetric vielbein condition between any pair of vielbeins is recovered for any superposition of N-vielbein interaction vertices in the absence of interaction cycles, we leave for future work, however. Here we conjecture that a matter coupling using an arbitrary linear superposition of building blocks \eqref{bbs-multi} of the form
\be \label{vielsup-multi}
\tilde E_{\rm eff} = \sum_i \alpha_i^{(1)} E_{(i)} + \sum_{i,j}^{i \neq j} \alpha_{i,j}^{(3)} {E_{(j)}} {E_{(i)}} E_{(j)}^{-1} + \sum_{i,j,k}^{i \neq j, j \neq k} \alpha_{i,j,k}^{(5)} {E_{(j)}} E_{(k)}^{-1} {E_{(i)}}  {E_{(k)}} E_{(j)}^{-1} + \ldots
\ee
will lead to a multi-gravity theory with a healthy $\Lambda_3$ decoupling limit (for subtleties in defining such a limit and why it is no longer unique in multi-gravity, see \cite{Noller:2013yja}). $i,j,k,\ldots$ here label all the $N$ vielbeins.

\section{Mini-superspace conditions for rank-2, -0 and mixed constructions} 
\label{appendix:msssumrules}

In Section~\ref{sec3:matterloops} we motivated splitting matter couplings in rank-2, rank-0 and mixed constructions. Here, we quantitatively show that these three classes are solutions to the mini-superspace condition \eqref{eqn:vaccond}. Equation \eqref{eqn:vaccond} may be combined with the explicit series expansion \eqref{eqn:expmss} to give,
\begin{equation}
\begin{split}
\det \left( \tilde{E} \right)  &\Bigg[  \frac{ \left( \sum_n \alpha_n N^n \right)''}{\sum_n \alpha_n N^n}  + 2 (D-1) \frac{ \left( \sum_n \alpha_n N^n \right)' \sum_m \alpha_m'}{\sum_n \alpha_n N^m \sum_m \alpha_m}   \\
& \;\;\; + (D-1) \frac{\sum_n \alpha_n''}{\sum_n \alpha_n}    + (D-1) (D-2)  \left(  \frac{\sum \alpha_n'}{\sum_n \alpha_n} \right)^2 \Bigg] = 0 , 
\end{split}
\label{eqn:msscond2}
\end{equation}
where for simplicity we have suppressed\footnote{Alternatively, one could simply set $E_{(p \neq 1)} = \delta$ and not worry about suppressed terms.} the $m$ index and used primes to denote differentiation with respect to $N$. 
By virtue of the analyticity of $\tilde{E}$, the $\alpha_{n}$ also admit a convergent series expansion, 
\begin{equation}
\alpha_n = \sum_{p_n \in \mathbb{Z}} c_{p_n} N^{p_n} ,
\label{eqn:exp2}
\end{equation}
for some $c_{p_n} (M, a,b)$. Then (\ref{eqn:msscond2}) becomes,
\begin{equation}
\begin{split}
\det \left( \tilde{E} \right) \sum_{l, m, n,  p_l, q_m, r_n \in \mathbb{Z}} & c_{p_l} c_{q_m} c_{r_n} \Big[  r_n^2 + r_n (2n-1) + 2p_l r_n + p_l ( 2n - 1) \\
&  + (D-1) p_l^2 + (D-1)(D-2) p_l q_m + n ( n-1)  \Big] N^{p_l + q_m + r_n + n -2} = 0 .
\end{split}
\end{equation}
This has the trivial solution $\sqrt{- \tilde{g}} = 0$, but also the nontrivial solutions found using the linear independence of $N^{b-2}$ (for every integer $b$),  
\begin{equation}
\begin{split}
\sum_{p_l, r_n \in \mathbb{Z}}  & c_{p_l} c_{b - p_l - r_n - n} c_{r_n} \Big[  r_n^2 + r_n (2n-1) + p_l r_n \left( D^2 - 3 D + 4  \right) \\
&+  p_l \Big( b (D-1)(D-2)  - n D (D - 3 )  - 1 \Big)+  p_l^2 (D-1) (3-D)  + n ( n-1)  \Big] = 0 . 
\end{split}
\label{eqn:msscond3}
\end{equation}
This is a severe restriction on the terms which may appear in the expansions (\ref{eqn:expmss}) and (\ref{eqn:exp2}). All candidate couplings which we consider must satisfy this sum relation. 

There are a number of immediate solutions to (\ref{eqn:msscond3}). If every $\alpha$ is a constant, then the powers $p_n = 0$, and the condition becomes,
\begin{equation}
\alpha_n \;  n (n-1 ) = 0
\label{eqn:ncond}
\end{equation}
and so only terms for which the total $E_{(1)}$ valence is $0$ or $1$ are allowed in $\tilde{E}$. These are our \emph{rank-2 constructions}. 
Additionally, we may consider a single $n=0$ term in (\ref{eqn:expmss}). $\alpha_0^D$ is then of dRGT form, 
\begin{equation}
c_{p_n} = \left( \begin{array}{c}
1/D \\
p_n
 \end{array}   \right)  \;\; \implies  \;\; \alpha_0^D = {\cI}_{\text{dRGT}}.
\end{equation}
These are our \emph{rank-0 constructions}.
Finally, in $D=1$ or $D=2$, the polynomial in (\ref{eqn:msscond3}) has roots which are independent of $b$, and therefore we can construct solutions like, 
\begin{equation}
\tilde{E} = \alpha \left( E_{(1)} \right) ^n  , \;\; \alpha \sim \left(  E_{(1)} \right) ^{-n} .
\end{equation}
But otherwise, the coefficients $c_{p_n}$ must be carefully (highly non-trivially) tuned to allow cancellation of the sum (\ref{eqn:msscond3}) for all $b$. These are our \emph{mixed constructions}.

\let\addtocontents\oldaddtocontent


\bibliographystyle{JHEP}
\bibliography{mn-bib}

\end{document}